\documentstyle[preprint,aps,epsf]{revtex} 
\catcode`@=11 
\newcount\@tempcntc 
\def\@citex[#1]#2{\if@filesw\immediate\write\@auxout{\string\citation{#2}}\fi 
  \@tempcnta\z@\@tempcntb\m@ne\def\@citea{}\@cite{\@for\@citeb:=#2\do 
    {\@ifundefined 
       {b@\@citeb}{\@citeo\@tempcntb\m@ne\@citea\def\@citea{,}{\bf ?}\@warning 
       {Citation `\@citeb' on page \thepage \space undefined}}% 
    {\setbox\z@\hbox{\global\@tempcntc0\csname b@\@citeb\endcsname\relax}% 
     \ifnum\@tempcntc=\z@ \@citeo\@tempcntb\m@ne 
       \@citea\def\@citea{,}\hbox{\csname b@\@citeb\endcsname}% 
     \else 
      \advance\@tempcntb\@ne 
      \ifnum\@tempcntb=\@tempcntc 
      \else\advance\@tempcntb\m@ne\@citeo 
      \@tempcnta\@tempcntc\@tempcntb\@tempcntc\fi\fi}}\@citeo}{#1}} 
\def\@citeo{\ifnum\@tempcnta>\@tempcntb\else\@citea\def\@citea{,}% 
  \ifnum\@tempcnta=\@tempcntb\the\@tempcnta\else 
   {\advance\@tempcnta\@ne\ifnum\@tempcnta=\@tempcntb \else \def\@citea{--}\fi 
    \advance\@tempcnta\m@ne\the\@tempcnta\@citea\the\@tempcntb}\fi\fi} 
\catcode`@=12 
\newcommand{\pslash}{\mbox{$\not{\hspace{-0.8mm}p}$}}

\newcommand{\sslash}{\mbox{$\not{\hspace{-0.8mm}s}$}}

\newcommand{\nc}{\newcommand} 
 
\nc{\be}{\begin{equation}} 
\nc{\ee}{\end{equation}} 
\nc{\bea}{\begin{eqnarray}} 
\nc{\eea}{\end{eqnarray}} 
\nc{\dn}{\Delta _0} 
   
\begin{document} 

\begin{titlepage} 
 \begin{flushleft} 
   MZ-TH/95-21\\ 
   TECHNION-PH 97-02
 \end{flushleft} 
 \begin{center} 
 {\Large \bf Inclusive Semileptonic Decays of Polarized $\Lambda_b$ Baryons  
into Polarized $\tau$-Leptons}\\[2cm] 
{\bf \large
 S. Balk$^\ast$,  J.G. K\"orner$^1$\footnote{Supported in part by
the BMFT, FRG under contract 06MZ566}  and D. Pirjol$^2$} \\[.5cm]
$^1$Johannes Gutenberg-Universit\"at,
 Institut f\"ur Physik, Staudingerweg 7,\\
 D-55099 Mainz, Germany\\
$^2$Department of Physics, Technion-Israel Institute of Technology,
 Haifa 32000, Israel\\[2cm]
\normalsize 
 \begin{abstract}   
 We employ OPE techniques within HQET to calculate the  
 inclusive semileptonic decays of polarized $\Lambda_b$ baryons. Lepton  
mass effects are included which enables us to also discuss rates into polarized  
$\tau$-leptons. We present explicit results for the longitudinal
polarization of the $\tau$ in the $\Lambda_b$ rest frame as well
as in the $(\tau ^-, \bar \nu_\tau)$ c.m. frame. In both 
the $\Lambda_b$ rest frame and in the $(\tau ^-, \bar \nu_\tau)$ c.m. frame
we make use of novel calculational techniques which
considerably simplify the calculations. The transverse
polarization components of the $\tau$ are calculated in the
$(\tau ^-, \bar \nu_\tau)$ c.m. frame. 
We delineate how to measure the full set of 14 polarized
and unpolarized structure functions of the 
decay process by angular correlation measurements.
A set of observables are identified that allow one to isolate
the contributions of the two $O(1/m_b^2)$
nonperturbative matrix elements $K_b$ and $\epsilon_b$.

\end{abstract}   
\end{center} 
\end{titlepage} 
\section{Introduction} 

Large samples of the bottom baryon $\Lambda_b$ have been produced  
on the $Z_0$ at LEP and are expected to be produced in future colliders.  
Advances in microvertexing techniques have allowed for efficient means of 
$\Lambda_b$ identification. For example, when LEP was 
running on the $Z$ one had  
$\approx 2.2\times 10^5$ $b\bar b$  
pairs per $10^6$ $Z$-decays. Of these approximately  
10\% go into $\Lambda_b$ baryons of  
which again $\approx$ 20\% decay  
semileptonically ($e, \mu$ and $\tau$). Thus one can expect a sample of 4000  
inclusive semileptonic $\Lambda_b$ (or $\bar \Lambda_b$) decays for every  
$10^6$ $Z$-decays. 
Plans at the SLC call for altogether $3 \times 10^6$
produced $Z$'s. The quality of the
$\Lambda_b$ data from the SLC will even be better because
its small beam size provides for an excellent definition of the
$\Lambda_b$ production vertex.
 
The $\Lambda_b$'s produced on the $Z$-peak are expected to be quite  
strongly polarized \cite{close}. 
This calls for a consideration of polarization effects  
in the 
$\Lambda_b$ decays which could be used to determine the polarization of the 
$\Lambda_b$. Approximately 10\% of the total ($ e+\mu +\tau $)  
semileptonic decay sample   
have a $\tau$-lepton in the final state. The $\tau$-lepton is sufficiently  
heavy which necessitates the inclusion of lepton mass effects in the  
dynamical rate calculations, apart from pure phase space effects. 
We will study also the
polarization of the $\tau$-lepton which, because of mass effects, differs
from its naive $m_l=0$ limiting
value. The $\tau$-polarization 
can be experimentally determined from its subsequent decay distributions. 
 
In this paper we present all the necessary tools to calculate  
inclusive semileptonic decays of polarized $\Lambda_b$'s including
$\tau$-lepton polarization effects and spin-spin, spin-momentum
and momentum-momentum correlation effects. 
Our calculation makes  
use of the heavy quark effective theory (HQET) and the operator product 
expansion (OPE) method as applied to heavy hadron decays. There is some  
overlap of our work with earlier results on  
inclusive $\Lambda_b$ decays in 
\cite{bialas,kraemer,koernerschuler,zalewski1,zal2,zal3,hagiwara} 
 and the more recent paper by M. Gremm, G. K\"opp and L.M. Sehgal on 
polarization effects in inclusive semileptonic 
$\Lambda_b$-decays \cite{gremm}. At the technical level 
we differ from the analysis of the above papers in that we employ helicity  
techniques  
to derive compact forms for the differential decay distributions  
including polarization effects. Also, by using suitable phase space  
variables, the integration of the higher order terms in the operator  
product expansion become much simplified, i.e. there are no surface term  
contributions. 
 
In Sec.II we list our results on the hadronic matrix elements needed for  
the subsequent semileptonic rate calculations. As the techniques of  
deriving the hadronic matrix elements are quite standard by now we do not  
dwell much on theoretical background but immediately proceed to the final  
results which we list in terms of a set of spin-dependent invariant  
structure functions. In Sec.III the invariant structure functions are  
related to helicity structure functions which determine the complete angular  
structure of the polarized decay distributions involving the three helicity
angles of the decay process. We write down the full  
differential decay distribution of polarized $\Lambda_b$-decays into  
longitudinally polarized $\tau $-leptons. We give analytical and numerical  
results on decay distributions integrated over phase space. 
In Sec.IV we discuss the case of transversally polarized $\tau $-leptons.

Sec.V is dedicated to a calculation of the longitudinal polarization of
the $\tau$ in the
$\Lambda_b$ rest frame as well as of its azimuthally averaged
transverse polarization component in the plane spanned
by the $\tau$ and the polarization vector of the $\Lambda_b$.
The calculation is considerably simplified by extracting the rate
as the absorptive part of the appropriate one-loop
contribution to the $\bar \tau \Lambda_b \rightarrow \bar \tau \Lambda_b $
scattering amplitude. Sec.VI contains our conclusions. More
detailed results have been collected in two Appendices. In 
Appendix A we list results on $q_0$-integrated helicity
structure functions. In Appendix B we give fully integrated
analytic results for the seven structure functions that describe
the angular decay structure of unpolarized $\Lambda_b$ decays. 
 
\section{Hadronic Matrix Elements} 

The dynamics of the hadron-side transitions is embodied in the hadronic  
tensor $W^{\mu\nu}$ which is defined as 
\bea 
W^{\mu\nu}_j(q_0,q^2,s)&=&(2\pi)^3\sum_X \delta^4(p_1-q-p_x) 
 \langle \Lambda_b(p_1,s) \vert J^{\mu\dag}_j \vert X_j (p_x)\rangle 
 \langle X_j (p_x)\vert J^\nu_j \vert \Lambda_b(p_1,s) \rangle 
 \, , 
\eea 
where $J^\mu_j (j=c,u)$ is the hadronic current inducing $b\rightarrow c$  
and $b\rightarrow u$ transitions. The hadron tensor is a function of two  
kinematic variables which we choose as $q_0$ and $q^2$. Note that we are  
not summing over the $\Lambda_b$-spin such that the hadron tensor depends  
also on its spin four-vector $s$. The structure of the  
hadron tensor can then be represented by an expansion along a standard 
set of covariants \cite{man} ($v=p_1/m_{\Lambda_b}$) 
\bea 
W^{\mu \nu} 
&=&-g^{\mu \nu}W_1+v^\mu v^\nu W_2 
            -i\epsilon ^{\mu \nu \rho \sigma} v_\rho q_\sigma W_3 
            +q^\mu q^\nu W_4 +(q^\mu v^\nu+ v^\mu q^\nu) W_5 
\nonumber \\ 
 &&+ 
\left[-q.s \left[-g^{\mu \nu}G_1+v^\mu v^\nu G_2 
 -i\epsilon^{\mu \nu \alpha \beta}v_\alpha q_\beta G_3 
 +q^\mu q^\nu G_4 + (q^\mu v^\nu+q^\nu v^\mu)G_5 \right] \right. 
\nonumber \\ 
&& + (s^\mu v^\nu+s^\nu v^\mu)G_6 
   + (s^\mu q^\nu+s^\nu q^\mu)G_7 
\nonumber \\ 
&& \left. +i\epsilon^{\mu \nu \alpha \beta}v_\alpha s_\beta G_8 
    +i\epsilon^{\mu \nu \alpha \beta}q_\alpha s_\beta G_9 \right]   
\nonumber \\ 
&&    + (s^\mu v^\nu-s^\nu v^\mu)G_{10} 
   + (s^\mu q^\nu-s^\nu q^\mu)G_{11} 
\nonumber \\ 
&&  +(v^\mu\epsilon^{\nu \alpha \beta \gamma} q_\alpha v_\beta s_\gamma 
   +v^\nu\epsilon^{\mu \alpha \beta \gamma} q_\alpha v_\beta s_\gamma) 
   G_{12} 
\nonumber \\ 
&&  +(q^\mu\epsilon^{\nu \alpha \beta \gamma} q_\alpha v_\beta s_\gamma 
   +q^\nu\epsilon^{\mu \alpha \beta \gamma} q_\alpha v_\beta s_\gamma) 
   G_{13} 
\, . 
\label{eq:spinkovars} 
\eea 
The last four invariants $G_{10}, G_{11}, G_{12}$ and $G_{13}$ are so-called
T-odd invariants which are fed by CP-odd and/or imaginary part  
contributions. They are zero for Standard Model couplings when loop effects
are neglected. They will therefore  
be disregarded in the following. Note that the invariants 
$W_4, W_5, G_4, G_5, G_7, G_{11}$ and $G_{13}$ do not contribute to  
inclusive semileptonic decays in the zero lepton mass case. 
%, also only the  
%linear combination $\frac{v.q}{q^2}G_8+G_9$ contributes in the 
%zero lepton mass case.  
Since we are also considering decays involving  
the massive $\tau$-lepton we must keep the full set of invariants  
implied by Eq.(\ref{eq:spinkovars}). 
 
In order to compute the spin-independent structure functions 
$W_1, ..., W_5$ and the spin-dependent structure functions 
$G_1, ..., G_9$ we resort to the well-known OPE techniques in HQET. The  
requisite steps in this calculation are so well documented in the  
literature \cite{gremm,man,chay,bigi,shifman,mannel,Ko,balk,FaLi}
that we can forgo a description of the  
intermediate steps and immediately list the final result of the OPE  
analysis. The hadron tensor is obtained as the absorptive part of the  
forward matrix element ($W^{\mu \nu}=-\frac{1}{\pi}$Im $T^{\mu \nu}$) 
\bea 
T^{\mu\nu}(q_0,q^2)&=&-i\langle \Lambda_b(p_1,s)\vert 
\int\mbox{d}^4x\,e^{-iq\cdot x}
\mbox{T}J^{\mu\dag}(x) J^\nu(0)
\vert\Lambda_b(p_1,s)\rangle \,.
\eea 
The amplitude $T^{\mu \nu}$ is decomposed into 14 invariant form-factors
$T_i,S_i$ analogous to the $W_i,G_i$ in (\ref{eq:spinkovars}) and can be
computed by using HQET methods 
\cite{gremm,chay,bigi,shifman,man,mannel}.
Keeping terms up to $1/m_b^2$ one obtains 
\bea 
T_1&=&\frac{1}{2\dn}(m_b-v.q)(1+X_b)+\frac{2m_b}{3}(K_b+G_b) 
\left( \frac{-1}{2\dn}+ \frac{q^2-(v.q)^2}{\dn^2} \right) \nonumber \\ 
 && +\frac{m_b (K_b+G_b)}{2\dn} -\frac{m_b^2 G_b}{3\dn^2}(m_b-v.q) 
,\nonumber \\ 
T_2&=&\frac{m_b}{\dn}(1+X_b)+\frac{2m_b}{3}(K_b+G_b) 
     \left( \frac{1}{\dn}+\frac{2m_b v.q}{\dn^2}  \right) \nonumber \\ 
 && +\frac{m_b (K_b+G_b)}{\dn}+\frac{4m_b^2 K_b v.q}{3\dn^2} 
    +\frac{2 m_b^3 G_b}{3 \dn^2} , \nonumber \\ 
T_3&=&\frac{1}{2\dn}(1+X_b)-\frac{2m_b}{3}(K_b+G_b) 
     \frac{m_b-v.q}{\dn^2}+\frac{2 m_b^2 K_b}{3 \dn^2} 
    -\frac{m_b^2 G_b}{3 \dn^2} \nonumber , \\ 
T_4&=&\frac{4 m_b}{3 \dn^2}(K_b+G_b)\nonumber , \\ 
T_5&=&\frac{-1}{2\dn}(1+X_b)-\frac{2m_b}{3}(K_b+G_b) 
     \frac{2m_b+v.q}{\dn^2} + \frac{m_b^2 G_b}{3 \dn^2}, 
\nonumber  \\ 
S_1 &=& -\frac{1+\epsilon_b}{2\Delta_0} 
 -\frac{5m_b}{3\Delta_0^2}v.q K_b 
 +\frac{4m_b^2 K_b}{3\Delta_0^3}(q^2-(v.q)^2) \, , 
\nonumber \\ 
S_2 &=& \frac{4m_b^2 K_b}{3\Delta_0^2} \, , 
\nonumber \\ 
S_3 &=& \frac{2m_b K_b}{3\Delta_0^2} \, , 
\nonumber \\ 
S_4 &=& 0 \, , 
\nonumber \\ 
S_5 &=& \frac{-2m_b K_b}{3\Delta_0^2} \, , 
\nonumber \\ 
S_6 &=& -\frac{m_b(1+\epsilon_b)}{2\Delta_0} 
  -\frac{5m_b K_b}{6\Delta_0} 
  -\frac{5m_b^2}{3\Delta_0^2}v.q K_b 
  +\frac{4m_b^3 K_b}{3\Delta_0^3}(q^2-(v.q)^2)  \, , 
\nonumber \\ 
S_7 &=&\frac{1+\epsilon_b}{2\Delta_0} 
  +\frac{(2m_b+3v.q)m_b K_b}{3\Delta_0^2} 
  -\frac{4m_b^2 K_b}{3\Delta_0^3}(q^2-(v.q)^2) \, , 
\nonumber \\ 
S_8 &=& \frac{m_b(1+\epsilon_b)}{2\Delta_0} 
  +\frac{m_b K_b}{6\Delta_0} 
  +\frac{5m_b^2}{3\Delta_0^2}v.q K_b 
  -\frac{4m_b^3 K_b}{3\Delta_0^3}(q^2-(v.q)^2) \, , 
\nonumber \\ 
S_9 &=& -\frac{1+\epsilon_b}{2\Delta_0} 
    -\frac{(2m_b+3v.q)m_b K_b}{3\Delta_0^2} 
    +\frac{4m_b^2 K_b}{3\Delta_0^3}(q^2-(v.q)^2) \, , 
\label{eq:tisi} 
\eea 
where 
\bea 
X_b&=&\frac{-2(m_b-v.q)m_b (K_b+G_b)}{\dn} 
     -\frac{8m_b^2 K_b}{3 \dn^2}(q^2-(v.q)^2) 
     +\frac{2 m_b^2 K_b}{\dn} \, , 
\eea 
We use the notation of \cite{man}
throughout the paper.
$K_b$ is related to the mean kinetic energy of the heavy quark in the 
$\Lambda_b$ baryon 
\bea 
 K_b&=& - \sum_s 
\langle \Lambda_b(p,s) \vert  
\bar b_v(x)\frac{(iD)^2}{2m_b^2} b_v(x)  \vert \Lambda_b(p,s) \rangle   
 \label{eq:kbwert}=\frac{\mu_\pi^2}{2m_b^2}\approx 0.013,
\eea 
where we used $\mu_\pi^2\approx 0.6$ GeV$^2$ \cite{CDNP} and $m_b=4.8$ GeV.
The spin-dependent contribution $\epsilon_b$ 
is defined by 
\bea 
\langle \Lambda_b(p,s) \vert 
 \bar b \gamma^\lambda \gamma_5 b  \vert \Lambda_b(p,s) \rangle 
 &=& (1+\epsilon_b) s^\lambda \, . 
\eea 
with $\epsilon_b\approx -\frac{2}{3} K_b$.
%The numerical value of the kinetic energy parameter
%$K_b\approx 0.013$ in Eq.(\ref{eq:kbwert}) is taken from 
%\cite{man}. 
An estimate of the spin-dependent parameter $\epsilon_b$
has been given in \cite{falk} with the result
$\epsilon_b=-\frac{2}{3}K_b$, based on an assumption that the
contribution of terms arising from double insertions of the
chromomagnetic operator can be neglected. A zero recoil sum rule
analysis gives the constraint $\epsilon_b\leq -\frac{2}{3}K_b$
\cite{koernerpirjol} which puts the estimate of \cite{falk}
at the upper boundary of the constraint. We use the value of
 \cite{falk} keeping in mind that 
the numerical value of 
$\epsilon_b$ could be reduced
in more realistic calculations.

In order to make our presentation as complete 
as possible we have retained the chromomagnetic 
contribution proportional to $G_b$ in Eq.(\ref{eq:tisi})
although it is zero for the $\Lambda_b$ system, i.e. 
\bea
G_b&:=&
\sum_s
\langle \Lambda_b(p,s) \vert
\bar b_v(x) 
\left(
 \frac{-gF_{\alpha \beta}\sigma^{\alpha \beta}}{4m_b^2}
\right)
 b_v(x)
\vert \Lambda_b(p,s)\rangle =0 \mbox{ . }
\label{eq:gb}
\eea
The reason is that the general helicity formalism introduced
later on can also be applied to $B$ meson and $\Omega_b$ baryon
decays where $G_b\neq 0$.

The denominator factor $\Delta_0$ is given by 
\bea 
 \Delta_0&=& (m_b v-q)^2-m_j^2+i\epsilon  
\eea 
The imaginary parts of inverse powers of $\Delta_0$, which are needed for
obtaining the structure functions $W_i,G_i$, can be obtained with the help 
of 
\bea 
\mbox{Im }\left(\frac{1}{\dn }\right) &=& 
\frac{-\pi}{2m_b} 
\delta \left( q_0- \left( \frac{-m_j^2+q^2+m_b^2}{2m_b}\right)\right)\,, 
\nonumber \\ 
\mbox{Im }\left(\frac{1}{\dn ^2}\right) &=& 
 \frac{-\pi}{4m_b^2} 
 \delta '\left( q_0-\left( \frac{-m_j^2+q^2+m_b^2}{2m_b}\right) \right)\,, 
\nonumber \\ \label{9}
\mbox{Im }\left(\frac{1}{\dn ^3}\right) &=& 
 \frac{-\pi}{16m_b^3} 
 \delta ''\left( q_0-\left( \frac{-m_j^2+q^2+m_b^2}{2m_b}\right) \right) \,. 
\eea 
Various subsets of (\ref{eq:tisi}) have appeared in the literature
 before \cite{gremm,man,balk}.
We have recalculated them and collected them together for ease of reference.

\section{Helicity structure functions and angular decay distributions} 

The hadronic structure for the transitions $\Lambda_b (s)\rightarrow X_j$ is  
fully specified by the absorptive parts of the 14 structure functions  
listed in Eq. (\ref{eq:tisi}). In order to obtain the full decay distribution  
for the inclusive decay 
$\Lambda_b (s)\rightarrow X_j+l^-(s_l)+\bar \nu_l$ one needs to contract  
the hadronic tensor $W^{\mu\nu}$ with the known leptonic tensor 
$L_{\mu\nu}$. Traditionally the contraction $L_{\mu\nu} W^{\mu\nu}$ 
is done in covariant fashion. Here we advocate a different approach and use  
helicity techniques to write down the relevant decay  
distributions. The advantage is that the angular decay distributions 
involving helicity angles are given 
by simple linear combinations of the helicity structure functions. 
The use of helicity techniques to describe angular decay distributions in  
exclusive semileptonic decays is widespread by now 
\cite{bialas,kraemer,koernerschuler} 
and is easily generalized to inclusive semileptonic decays. 
 
Let us begin by writing down the relation between the full spin-dependent  
 set of 14 helicity structure functions and the set of 14 invariant 
structure functions in Eq. (\ref{eq:spinkovars}). One has 
\bea 
W_{++}^{++}&=& (q_0^2-q^2)G_3 - (G_1 + W_3)p - q_0 G_9 - G_8 + W_1 \, , 
\nonumber\\ 
W_{++}^{--}&=& -((q_0^2-q^2)G_3 - (G_1 - W_3)p - q_0 G_9 - G_8 - W_1) \, , 
\nonumber\\ 
W_{--}^{++}&=& -((q_0^2-q^2)G_3 + (G_1 - W_3)p - q_0 G_9 - G_8 - W_1) \, , 
\nonumber\\ 
W_{--}^{--}&=& (q_0^2-q^2)G_3 + (G_1 + W_3)p - q_0 G_9 - G_8 + W_1 \, , 
\nonumber\\ 
W_{tt}^{++}&=&\frac{( - ((2(q^2 G_5 - G_6)q_0 - (G_1 + 2 G_7) q^2 + q_0^2  
G_2 
                 + q^4 G_4)p - q_0^2 W_2 - 2 q_0 q^2 W_5 - q^4 W_4 
                + q^2 W_1))}{q^2} \, , 
\nonumber\\ 
W_{tt}^{--}&=&\frac{((2(q^2 G_5 - G_6)q_0 - (G_1 + 2 G_7) q^2 + q_0^2 G_2 
                 + q^4 G_4)p + q_0^2 W_2 + 2 q_0 q^2 W_5 + q^4 W_4 
                - q^2 W_1)}{q^2} \, , 
\nonumber\\ 
W_{00}^{++}&=& \frac{(q_0^2  - q^2) W_2 + (2 q_0 G_6 - q^2 G_1) p - 
 p^3 G_2 + q^2 W_1}{q^2} \, , 
\nonumber\\ 
W_{00}^{--}&=& \frac{(q_0^2  - q^2) W_2 - (2 q_0 G_6 - q^2 G_1) p + 
 p^3 G_2 + q^2 W_1}{q^2} \, , 
\nonumber\\ 
W_{0t}^{++}&=&\frac{p q_0 W_2 + p q^2 W_5 - q_0^3 G_2 
              - q_0^2 q^2 G_5 + 2 q_0^2 G_6 
            + q_0 q^2 G_2 + q_0 q^2 G_7 + q^4 G_5 - q^2 G_6}{q^2} \, , 
\nonumber\\ 
W_{0t}^{--}&=&\frac{p q_0 W_2 + p q^2 W_5 + q_0^3 G_2 
              + q_0^2 q^2 G_5 - 2 q_0^2 G_6 
            - q_0 q^2 G_2 - q_0 q^2 G_7 - q^4 G_5 + q^2 G_6}{q^2} \, , 
\nonumber\\ 
W_{0+}^{+-}&=&\frac{-2 (p G_6 + q_0 G_8 + q^2 G_9)}{\sqrt{q^2}\sqrt{2}} \,  
, 
\nonumber\\ 
W_{t+}^{+-}&=&\frac{-2 (p G_8 + q_0 G_6 + q^2 G_7)}{\sqrt{q^2}\sqrt{2}} \,  
, 
\nonumber\\ 
W_{-t}^{+-}&=&\frac{-2 (p G_8 - q_0 G_6 - q^2 G_7)}{\sqrt{q^2}\sqrt{2}} \,  
, 
\nonumber\\ 
W_{-0}^{+-}&=&\frac{2 (p G_6 - q_0 G_8 - q^2 G_9)}{\sqrt{q^2}\sqrt{2}} \, .
\label{eq:helamplituden} 
\eea 
We denoted here $p=\sqrt{q_0^2-q^2}$. The helicity structure functions  
$W_{\lambda_W \lambda'_W}^{\lambda_{\Lambda_b} \lambda'_{\Lambda_b}}$ 
are defined by 
\bea 
W_{\lambda_W \lambda'_W}^{\lambda_{\Lambda_b} \lambda'_{\Lambda_b}} 
&=& (2\pi)^3
\sum_X \delta^4(p_1-q-p_x) 
 \langle X_j \vert J^\mu_j \vert \Lambda_b , \lambda_{\Lambda_b} \rangle 
\epsilon_\mu^\ast (\lambda_W) 
 \langle \Lambda_b, \lambda'_{\Lambda_b} \vert J^{\nu\dag}_j  
\vert X_j \rangle 
\epsilon_\nu (\lambda'_W) \,.
\label{eq:helis} 
\eea 
Here $\lambda_{\Lambda_b}=\pm 1/2$ is the helicity of the $\Lambda_b$
and $\lambda_W=0,\pm 1,t$ are the helicities of the virtual $W$-boson
(spatial: $\lambda_W=0,\pm 1$; temporal: $\lambda_W=t$).
Note that there are no zeroth order parton model and kinetic
energy contributions to the structure functions $W_{++}^{++}$
and $W_{--}^{--}$ because of angular momentum conservation.
We shall return to this point later in this section 
when we discuss the contribution
of the spin-dependent matrix element $\epsilon_b$.
Note that from Eq. (\ref{eq:helis}) one has the hermiticity relation 
\bea 
  W_{\lambda_W \lambda '_W}^{\lambda_{\Lambda_b}\lambda'_{\Lambda_b}\ast} 
  &=& W_{\lambda '_W \lambda_W}^{\lambda'_{\Lambda_b}\lambda_{\Lambda_b}} 
 \, \mbox{.} 
 \label{eq:wwsym} 
\eea 
Since the helicity structure functions are real in our case one can drop the  
complex conjugation sign in Eq.(\ref{eq:wwsym}). The helicity structure  
functions are defined in the $\Lambda_b$ rest system. We therefore need to  
specify a $z$-axis which we take to be along $\vec p_X$ (see Fig.1). 

\begin{figure}
\begin{center}
$\quad$\\
\vspace*{-0cm}
\hspace*{-0cm}
\begin{picture}(0,0)%
\includegraphics{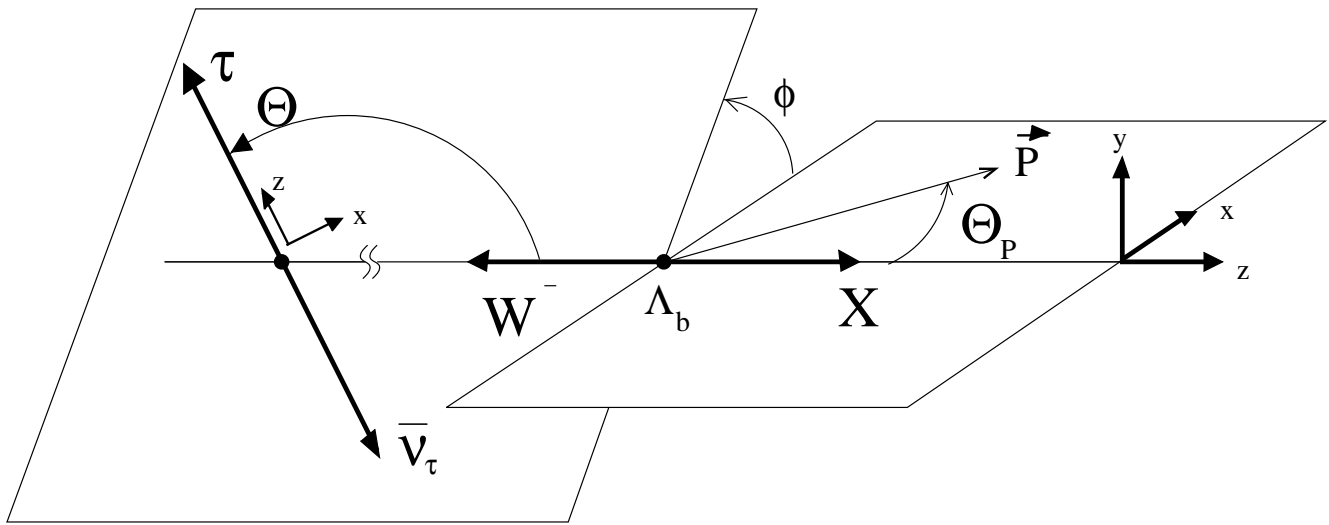}%
\end{picture}%
\setlength{\unitlength}{0.012500in}%
\begin{picture}(552,215)(55,610)
\end{picture}
\end{center}
\caption{Definition of the polar angles $\Theta$ and $\Theta_P$ and of
the azimuthal angle $\phi$
in the decay $\bar \Lambda_b \to X$ $+ W^-( \to l^- \bar \nu _l )$ 
in the $\Lambda_b$ rest system. We 
specify a $z$-axis which we take to be along $\vec p_X$.
$\vec P$ denotes the polarization three-vector of the $\Lambda_b$.}
\label{dichtwinkel}
\end{figure}

As noted before the full angular decay distribution of 
$\Lambda_b (s)\rightarrow X_j+l^-(s_l)+\bar \nu_l$ including all 
polarization effects is completely determined by the set of 14 helicity
structure functions. The necessary manipulations  
involving Wigner's $D^J_{m m'}$ functions are standard and well documented  
in the literature \cite{rose,itzykson} (see also Sec.IV).
Here we closely follow the presentation of 
\cite{zalewski1,zal2,zal3}.
 For example,  
for the five-fold decay distribution in  
$q_0, q^2, \cos\Theta, \cos \Theta_P$ and $\phi$
into negative (d$\Gamma^-$) and positive (d$\Gamma^+$) helicity leptons
we obtain 
\footnote{Similar decay distributions have been written down in
 \cite{lampe}, where the $O(\alpha_s)$ corrections to unpolarized
 $t\rightarrow b$ decays were evaluated.}
\bea 
&&\frac{\mbox{d}\Gamma^-}{\mbox{d}q_0 \mbox{d}q^2 
\mbox{d}\cos\Theta \mbox{d}\cos \Theta_P \mbox{d}\phi} 
=\frac{2 G^2\vert V_{bj} \vert ^2 (q^2-m_l^2)^2 \sqrt{q_0^2-q^2}}{3 (2\pi)^4  
q^2} 
\nonumber \\ 
&&\left[ \left( \rho_{++}\left( 
                  W_{--}^{++}  +W_{++}^{++}\right) 
           +\rho_{--}\left( 
                  W_{--}^{--}  +W_{++}^{--}\right) \right) 
  \frac{3}{8}(1+\cos^2\Theta )\right. 
\nonumber \\ 
&& 
+\left( \rho_{++} W_{00}^{++}+\rho_{--} W_{00}^{--}\right) 
 \frac{3}{4} \sin^2\Theta  
\nonumber \\ 
&& 
+\frac{3}{4}\left(\rho_{++}\left( 
      W_{++}^{++}-W_{--}^{++} \right)  
    +\rho_{--}\left(W_{++}^{--}-W_{--}^{--}\right)\right) \cos\Theta 
\nonumber \\ 
&& 
+\frac{3}{2 \sqrt{2}} \rho_{+-}\left( W_{-0}^{+-}+W_{0+}^{+-}\right) 
 \sin\Theta \cos\phi 
\nonumber \\ 
&& 
\left. 
+\frac{3}{4 \sqrt{2}} \rho_{+-}\left( W_{-0}^{+-}-W_{0+}^{+-}\right) 
 \sin 2\Theta \cos\phi \right] 
\label{eq:gminus} 
\\ 
&&\frac{\mbox{d}\Gamma^+}{\mbox{d}q_0 \mbox{d}q^2 
\mbox{d}\cos\Theta \mbox{d}\cos \Theta_P \mbox{d}\phi} 
=\frac{2 G^2\vert V_{bj} \vert ^2 (q^2-m_l^2)^2 \sqrt{q_0^2-q^2}}{3 (2\pi)^4  
q^2}  
\nonumber \\ 
&&\frac{m_l^2}{2q^2}\left[ \left( \rho_{++}\left( 
                  W_{--}^{++}  +W_{++}^{++}\right) 
           +\rho_{--}\left( 
                  W_{--}^{--}  +W_{++}^{--}\right) \right) 
  \frac{3}{4}\sin^2\Theta \right. 
\nonumber \\ 
&& 
\mbox{\hspace*{1cm}} 
+\left( \rho_{++} W_{00}^{++}+\rho_{--} W_{00}^{--}\right) 
 \frac{3}{2} \cos^2\Theta  
\nonumber \\ 
&& 
\mbox{\hspace*{1cm}} 
+\frac{3}{2}\left(\rho_{++} 
      W_{tt}^{++} 
    +\rho_{--} W_{tt}^{--}\right) 
\nonumber \\ 
&& 
\mbox{\hspace*{1cm}} 
+3\left(\rho_{++} 
      W_{0t}^{++} 
    +\rho_{--} W_{0t}^{--}\right) \cos\Theta  
\nonumber \\ 
&& 
\mbox{\hspace*{1cm}} 
-\frac{3}{2} \sqrt{2} \rho_{+-}\left( W_{-t}^{+-}-W_{t+}^{+-}\right) 
 \sin\Theta \cos\phi 
\nonumber \\ 
&& 
\left. 
\mbox{\hspace*{1cm}} 
-\frac{3}{4} \sqrt{2} \rho_{+-}\left( W_{-0}^{+-}-W_{0+}^{+-}\right) 
 \sin 2\Theta \cos\phi \right] 
\label{eq:gplus} 
\eea 
For $\Lambda_c \rightarrow X_s + l^+ +\nu_l$ decays 
one has to effect the replacement d$\Gamma^\mp \leftrightarrow 
\mbox{d}\Gamma^\pm$
and one has to change the signs of the
parity violating contributions proportional to $\cos \Theta$ and
$\sin \Theta \sin \phi$ in Eq.(\ref{eq:gminus}).
The differential rate into unpolarized leptons is simply
d$\Gamma^++$d$\Gamma^-$.
The polar angles $\Theta$ and $\Theta_P$ and the azimuthal angle 
$\phi$ are defined in Fig.\ref{dichtwinkel}. 
 We have rotated the density matrix of  
the $\Lambda_b$ to the $z$-axis such that one has 
\bea 
 \rho^{\Lambda_b}(\cos\Theta_P)&=& 
 \frac{1}{2}\left( \begin{array}{cc} 
 1+P \cos\Theta_P & P \sin\Theta_P\\ 
 P \sin\Theta_P & 1-P\cos\Theta_P 
 \end{array}
 \right) \, . 
 \label{eq:dichteml} 
\eea 
Note that for  
$\Lambda_b$'s from $Z$-decays one expects 
that the $\Lambda_b$'s are longitudinally polarized with backward  
polarization. Thus, for $\Lambda_b$'s from $Z$-decays, the direction of $P$  
in Fig.\ref{dichtwinkel} coincides with the boost direction that brings  
$\Lambda_b$ to rest ($P\geq 0$). 

The longitudinal polarization of the  
$\tau$-lepton is given by 
\bea 
P_{\tau}^l&=& \frac{d\Gamma^+ -d\Gamma^-}{d\Gamma^+ +d\Gamma^-} 
\label{eq:longipol} 
\eea 
We emphasize that the longitudinal polarization of the $\tau$ calculated from 
(\ref{eq:gminus}, \ref{eq:gplus}) and (\ref{eq:longipol}) refers to the 
$(\tau, \bar \nu_\tau)$ c.m. frame which differs from the longitudinal  
polarization of the $\tau$ in the $\Lambda_b$ rest system as calculated 
e.g. in \cite{gremm,FaLi}.
A slight adaptation of the lepton-side density matrix elements in  
(\ref{eq:gminus}, \ref{eq:gplus}) 
as described in \cite{hagiwara} will yield the longitudinal 
polarization in the $\Lambda_b$ rest system. 
Put in a different language the two respective polarizations 
are related to one another by a Wigner rotation (see e.g. \cite{vives}).
A direct computation of the longitudinal polarization of the
$\tau$ in the $\Lambda_b$ rest frame including correlation effects
will be presented in Sec.V.
 
The quasi three-body decay 
$\Lambda_b (s)\rightarrow X_j+l^-+\bar \nu_l$ 
is described by three kinematic invariants. A particularly convenient  
choice is the one used in Eqs.(\ref{eq:gminus}, \ref{eq:gplus}) 
 in terms of $q_0, q^2$ and $\cos\Theta$.
Working in $(q_0, q^2, \cos\Theta)$ phase space has big technical
advantages as can be seen in the following.
The integration over $\cos\Theta$ is trivial. The integration over $q_0$
is almost trivial as can be seen by the following reasoning.  
The leading order parton model or  
free quark decay contributions are proportional to the $\delta$-function  
and the $q_0$-integration amounts to the substitution 
$q_0=\frac{m_b^2-m_j^2+q^2}{2m_b}$ in these terms. The corrections to the  
parton model contributions involve derivatives of the  
$\delta$-function. Using partial integration the derivatives can easily be  
shifted to the integrand functions without encountering surface term  
contributions because, to the requisite order in $1/m_b$, the two-dimensional
parton phase space never touches the boundary of the three-dimensional  
particle phase space, except at maximal $q^2$ and $q_0$ where the  
integration measure is zero. The only nontrivial phase-space integration  
that remains to be done is with regard to $q^2$. 
  
If desired, the transformation to the usual
$(q_0, q^2, E_l)$ set of variables can be done
with the help of the relation 
\bea 
\cos\Theta &=& \frac{q_0(q^2+m_l^2)-2q^2E_l}{\sqrt{q_0^2-q^2}(q^2-m_l^2)}\,.
\label{eq:cositet}  
\eea 
%implying 
%\bea 
%d\cos\Theta &=& -\frac{2q^2}{\sqrt{q_0^2-q^2}(q^2-m_l^2)}dE_l\,.
%\label{eq:dcositet}  
%\eea 
However, as can be easily appreciated by substituting  
Eqs.(\ref{eq:cositet})
%and (\ref{eq:dcositet}) 
in the differential rate  
Eq.(\ref{eq:gminus}, \ref{eq:gplus}), the integration of the corresponding 
differential rate function is more cumbersome for the following two reasons.
First, when doing the $q_0$-integration, one has to carefully consider the
contributions from the surface terms induced by the derivatives of the
$\delta$-function from the OPE expansion (see Eqs.(\ref{9})). These may
lead to the appearance of spurious singularities when calculating 
polarization type observables \cite{GrLi,gremm}.
Second, one remains with two nontrivial ($E_l$ and $q^2$) integrations,
apart from having to carefully consider surface term contributions when 
doing the $q_0$-integration.

\begin{center}
\begin{tabular}{|c|c|c|c|c|}  
\hline
%\rule{0pt}{11.5pt}
& \multicolumn{2}{|c|}{$b\rightarrow c$} & \multicolumn{2}{c|}{$b\rightarrow u$} \\ 
\cline{2-5}
& $\eta=m_\tau^2/m_b^2$ & $\eta=0$ & $\eta=m_\tau^2/m_b^2$ & $\eta=0$ \\ 
\hline 
\hline
$\hat\Gamma_U^-$ 
& 0.0389 & 0.177 & 0.133 & 0.352\\ 
& (0.0376) & (0.172) & (0.124) & (0.333)\\  
\hline 
$\hat\Gamma_L^-$
& 0.0330 & 0.332 & 0.124 & 0.635\\ 
\,& (0.0352) & (0.344) & (0.136) & (0.666)  \\ \hline 
$\hat\Gamma_F^-$
& --0.0185 & --0.105 & --0.093 & --0.282 \\ 
\,& (--0.0208)  & (--0.112)  & (--0.124)  & (--0.333) \\ \hline 
$\hat\Gamma_U^{P-}$ 
& 0.0183 & 0.104 & 0.092 & 0.279 \\ 
\,& (0.0208) & (0.112) & (0.124) & (0.333) \\ \hline 
$\hat\Gamma_L^{P-}$ 
& --0.0282 & --0.322 & --0.116 & --0.629 \\ 
\,& (--0.0306) & (--0.331) & (--0.136) & (--0.666) \\ \hline 
$\hat\Gamma_F^{P-}$ 
& --0.0385 & --0.176 & --0.132 & --0.349\\ 
\,& (--0.0376) & (--0.172) & (--0.124) & (--0.333) \\ \hline 
$\hat\Gamma_I^{P-}$ 
& --0.0397 & --0.208 & --0.123 & --0.418 \\ 
\,& (--0.0349) & (--0.213) & (--0.128) & (--0.431) \\ \hline 
$\hat\Gamma_A^{P-}$ 
& --0.0239 & --0.179 & --0.103 & --0.390 \\ 
\,& (--0.0263) & (--0.186) & (--0.128) & (--0.431) \\ \hline 
\end{tabular}
\end{center}
\begin{quote} {\bf Table 1.}
Numerical values for polarized and unpolarized helicity
structure functions into negative helicity leptons. Column 2: 
$b\rightarrow c, l=\tau$; column 3: $b\rightarrow c, l=e$ ($m_e=0$);
 column 4: $b\rightarrow u $ $(m_u=0)$, $l=\tau$;
 column 5: $b\rightarrow u, l=e$ ($m_e=0$).
\end{quote}

Returning to the differential rate Eq.(\ref{eq:gminus}, \ref{eq:gplus}) one
first does the $q_0$-integration which, after integration by parts,
 amounts to a mere substitution as noted before. 
The results of the $q_0$-integration are listed in Appendix A.
Next one integrates over $q^2$ in the limits
$m_l^2\leq q^2\leq (m_1-m_2)^2$.
Numerical and analytical results of the $q^2$-integration
can be found in Tables 1 and 2 and in Appendix B.
 
\begin{center} 
\begin{tabular}{|c|c|c|} 
\hline  
& $\qquad b\rightarrow c\qquad$ & $\qquad b\rightarrow u\qquad$ \\ 
\hline 
\hline
$\hat\Gamma_U^+$ 
&0.0081 & 0.0198 \\ 
&(0.0079) & (0.0186) \\ 
\hline 
$\hat\Gamma_L^+$ 
&0.0076 &0.0218 \\ 
&(0.0080) &(0.0234) \\ 
\hline 
$\hat\Gamma_S^+$ 
&0.0079 &0.0231 \\ 
&(0.0080) &(0.0234) \\ 
\hline 
$\hat\Gamma_{SL}^+$ 
&0.0066 &0.0210 \\ 
&(0.0072) &(0.0234) \\ 
\hline 
$\hat\Gamma_U^{P+}$
&0.0041 &0.0154 \\ 
&(0.0046) &(0.0186) \\ 
\hline 
$\hat\Gamma_L^{P+}$
&-0.0067 &-0.0213 \\ 
&(-0.0072) &(-0.0234) \\ 
\hline 
$\hat\Gamma_S^{P+}$
&-0.0066 &-0.0208 \\ 
&(-0.0072) &(-0.0234) \\ 
\hline 
$\hat\Gamma_{SL}^{P+}$ 
&-0.0077 &-0.0222 \\ 
&(-0.0080) & (-0.0234) \\ 
\hline 
$\hat\Gamma_{ST}^{P+}$ 
&-0.0077 &-0.0209 \\ 
&(-0.0076)  & (-0.0205) \\ 
\hline 
$\hat\Gamma_A^{P+}$
&-0.0055 &-0.0181 \\ 
&(-0.0060) &(-0.0205) \\ 
\hline 
\end{tabular}
\end{center}
\begin{quote}
{\bf Table 2.} Numerical values for polarized and unpolarized
helicity structure functions into positive helicity $\tau$-leptons
for $b\rightarrow c$ and $b\rightarrow u$ transitions. 
Parameter values as in Table 1.
\end{quote}

We use in the Tables 1,2 the following numerical values:
$m_\tau=1.777$ GeV,
$m_c=1.451$ GeV,  
$m_b=4.808$ GeV. $K_b$ is given by
$K_b=\mu_\pi^2/(2 m_b^2)$ with $\mu_\pi^2=0.6$ GeV$^2$
and for $\epsilon_b$ we adopt the value $\epsilon_b=-2K_b/3$ as explained
above. The values in brackets show the free quark decay (FQD) results.

In order to simplify our notation we introduce a set of unpolarized and  
polarized reduced differential rate functions $d\hat\Gamma_i^-$ and  
$d\hat\Gamma_i^{P-}$, respectively, for the decays into negative helicity 
leptons. We define scaled variables  
$\hat q^2=q^2/m_b^2$, $\hat q_0=q_0/m_b$, $\rho=m_j^2/m_b^2$ and 
$\eta=m_\tau^2/m_b^2$  
and write   
\bea 
\frac{\mbox{d}\hat\Gamma_U^{(P)-}}{\mbox{d}\hat q^2} 
&=& 16
\frac{(\hat q^2-\eta)^2}{\hat q^2} 
I(W_{++}^{++}+W_{--}^{++} +(-) (W_{++}^{--}+W_{--}^{--})) 
\nonumber\\ 
\frac{\mbox{d}\hat\Gamma_L^{(P)-}}{\mbox{d}\hat q^2} 
&=& 16
\frac{(\hat q^2-\eta)^2}{\hat q^2} 
I(W_{00}^{++}+(-) W_{00}^{--}) 
\nonumber\\ 
\frac{\mbox{d}\hat\Gamma_F^{(P)-}}{\mbox{d}\hat q^2} 
&=& 16
\frac{(\hat q^2-\eta)^2}{\hat q^2} 
I(W_{++}^{++}-W_{--}^{++} +(-) (W_{++}^{--}-W_{--}^{--})) 
\nonumber\\ 
\frac{\mbox{d}\hat\Gamma_{I}^{P -}}{\mbox{d}\hat q^2} 
&=& 16
\frac{(\hat q^2-\eta)^2}{\hat q^2} 
I(W_{-0}^{+-}+ W_{0+}^{+-}) 
\nonumber\\ 
\frac{\mbox{d}\hat\Gamma_{A}^{P -}}{\mbox{d}\hat q^2} 
&=& 16
\frac{(\hat q^2-\eta)^2}{\hat q^2} 
I(W_{-0}^{+-}- W_{0+}^{+-}) 
\label{eq:upm}
\eea 
where d$\hat\Gamma_U^{(P)-}$ stands for either d$\hat\Gamma_U^{-}$
or d$\hat\Gamma_U^{P-}$ etc. with the corresponding signs 
specified on the r.h.s. of Eq. (\ref{eq:upm}).
Accordingly we define reduced differential rate functions for decays into  
positive helicity leptons. 
\bea 
\frac{\mbox{d}\hat\Gamma_{U}^{(P)+}}{\mbox{d}\hat q^2} 
&=& 
\frac{\eta}{2 \hat q^2} 
\frac{\mbox{d}\hat\Gamma_{U}^{(P)-}}{\mbox{d}\hat q^2} 
\nonumber\\ 
\frac{\mbox{d}\hat\Gamma_{L}^{(P)+}}{\mbox{d}\hat q^2} 
&=& 
\frac{\eta}{2 \hat q^2} 
\frac{\mbox{d}\hat\Gamma_{L}^{(P)-}}{\mbox{d}\hat q^2} 
\nonumber\\ 
\frac{\mbox{d}\hat\Gamma_{S}^{(P)+}}{\mbox{d}\hat q^2} 
&=& 16
\frac{\eta}{2 \hat q^2} 
\frac{(\hat q^2-\eta)^2}{\hat q^2} 
I(W_{tt}^{++} +(-) W_{tt}^{--}) 
\nonumber\\ 
\frac{\mbox{d}\hat\Gamma_{SL}^{(P)+}}{\mbox{d}\hat q^2} 
&=& 16
\frac{\eta}{2 \hat q^2} 
\frac{(\hat q^2-\eta)^2}{\hat q^2} 
I(W_{0t}^{++} +(-) W_{0t}^{--}) 
\nonumber\\ 
\frac{\mbox{d}\hat\Gamma_{ST}^{P +}}{\mbox{d}\hat q^2} 
&=& 16
\frac{\eta}{2 \hat q^2} 
\frac{(\hat q^2-\eta)^2}{\hat q^2} 
I(W_{-t}^{+-} - W_{t+}^{+-}) 
\nonumber\\ 
\frac{\mbox{d}\hat\Gamma_{A}^{P +}}{\mbox{d}\hat q^2} 
&=& 
\frac{\eta}{2 \hat q^2} 
\frac{\mbox{d}\hat\Gamma_A^{P -}}{\mbox{d}\hat q^2}
 \label{eq:upp}
\eea 
Note that the structure function combination
$I(W_{-t}^{+-} + W_{t+}^{+-}) $ does not appear in Eqs. (\ref{eq:upm})
and (\ref{eq:upp}). This combination can only be measured
through the transverse spin components of the $\tau$-lepton
as discussed in the next section.
The differential rate into negative helicity leptons can then be written as 
\bea 
&&\frac{\mbox{d}\Gamma^-}{\mbox{d}\hat q^2 \mbox{d}\cos
\Theta \mbox{d}\cos \Theta_P \mbox{d}\phi} 
=\frac{\Gamma_b}{4\pi}\times\nonumber \\ 
&&\left[ \frac{3}{8} \left(  
\frac{d\hat\Gamma_U^-}{d\hat q^2} 
+P\cos\Theta_P \frac{d\hat\Gamma_U^{P -}}{d\hat q^2} 
\right)(1+\cos^2\Theta )\right. 
\nonumber \\ 
&& 
+ \frac{3}{4} \left(  
\frac{d\hat\Gamma_L^-}{d\hat q^2} 
+P\cos\Theta_P \frac{d\hat\Gamma_L^{P -}}{d\hat q^2} 
\right)\sin^2\Theta  
\nonumber \\ 
&& 
+ \frac{3}{4} \left(  
\frac{d\hat\Gamma_F^-}{d\hat q^2} 
+P\cos\Theta_P \frac{d\hat\Gamma_F^{P -}}{d\hat q^2} 
\right)\cos\Theta  
\nonumber \\ 
&& 
\left. 
+ \frac{3}{2 \sqrt{2}} \frac{d\hat\Gamma_I^{P -}}{d\hat q^2} 
P\sin\Theta_P \sin\Theta\cos\phi 
- \frac{3}{4 \sqrt{2}} \frac{d\hat\Gamma_A^{P -}}{d\hat q^2} 
P\sin\Theta_P\sin 2\Theta\cos\phi 
\right] 
\label{eq:uppmm}
\eea 
where
\bea
 \Gamma_b &=& \frac{\vert V_{bj}\vert ^2 G_F^2 m_b^5}{192 \pi^3} 
 \mbox{ .}
\eea
For the rate into positive helicity leptons we have 
\bea 
&&\frac{\mbox{d}\Gamma^+}{\mbox{d}\hat q^2 \mbox{d}\cos\Theta 
\mbox{d}\cos \Theta_P \mbox{d}\phi} 
=\frac{\Gamma_b}{4\pi}\times\nonumber \\ 
&&\left[ \frac{3}{4} \left(  
\frac{d\hat\Gamma_U^+}{d\hat q^2} 
+P\cos\Theta_P \frac{d\hat\Gamma_U^{P +}}{d\hat q^2} 
\right)\sin^2\Theta \right. 
\nonumber \\ 
&& 
+ \frac{3}{2} \left(  
\frac{d\hat\Gamma_L^+}{d\hat q^2} 
+P\cos\Theta_P \frac{d\hat\Gamma_L^{P +}}{d\hat q^2} 
\right)\cos^2\Theta  
\nonumber \\ 
&& 
+ \frac{3}{2} \left(  
\frac{d\hat\Gamma_S^+}{d\hat q^2} 
+P\cos\Theta_P \frac{d\hat\Gamma_S^{P +}}{d\hat q^2} 
\right) 
\nonumber \\ 
&& 
+ 3 \left(  
\frac{d\hat\Gamma_{SL}^+}{d\hat q^2} 
+P\cos\Theta_P \frac{d\hat\Gamma_{SL}^{P +}}{d\hat q^2} 
\right)\cos\Theta  
\nonumber \\ 
&& 
\left. 
+ \frac{3}{2} \sqrt{2} \frac{d\hat\Gamma_{ST}^{P +}}{d\hat q^2} 
P\sin\Theta_P\sin\Theta\cos\phi 
+ \frac{3}{4} \sqrt{2} \frac{d\hat\Gamma_A^{P +}}{d\hat q^2} 
P\sin\Theta_P\sin 2\Theta\cos\phi 
\right] 
\eea 
For quick reference we shall adopt a generic
labelling for the various helicity structure functions, namely
we write $U^{(P)-}$ for d$\Gamma_U^{(P)-}/\mbox{d}\hat q^2$, etc.
An inspection of the $q^2$ dependence of the helicity structure 
functions shows that the structure functions 
$U^{(P)-}$, $L^{(P)-}$, $F^{(P)-}$, $U^{(P)+}$, $L^{(P)+}$,
  $S^{(P)+}$ and $SL^{(P)+}$ can be integrated analytically. In contrast
to this, the integrations of the structure functions 
$I^{(P)-}$, $A^{(P)-}$, $ST^{(P)+}$ and $A^{(P)+}$ lead to
incomplete elliptic functions and will therefore be performed
numerically.

In Appendix B we list analytic results for the totally integrated 
unpolarized structure functions 
$U^{-}$, $L^{-}$, $F^{-}$, $U^{+}$, $L^{+}$ and $S^{+}$ as well as for the
total rate function proportional to
$(U^{-}+ U^{+}+ L^{-} +L^{+}+3 S^{+})$. Numerical results are listed
in Tables 1 and 2, where we have used the following set of
input values: $m_b=4.8$ GeV, $m_c=1.45$ GeV, $m_u=0$ GeV,
 $m_\tau=1.777$ GeV, $K_b=\mu_\pi^2/(2m_b^2)$, $\mu_\pi^2=0.6$ GeV$^2$
 and $\epsilon_b=-2 K_b/3$.
In order to assess the importance of the nonperturbative contributions
we also list the zeroth order parton model values ($K_b=\epsilon_b=0$)
in brackets. Judging from the numerical entries in Tables 1 and
2 the effect of the nonperturbative contributions can go both ways,
i.e. depending on the particular structure function, they can 
enhance or decrease the parton model results. The nonperturbative
contributions are generally
small, at the few percent level. Notable are the
structure functions $U^{P-}$, $A^{P-}$, $SL^{+}$,
  $U^{P+}$ and $L^{P+}$ where the nonperturbative contributions
exceed 10\%. 

Fully integrated values of the longitudinal polarization
of the $\tau$ can easily be constructed in analogy to 
Eq. (\ref{eq:longipol}). It is also evident that there are
no transitions into positive helicities for $m_l=0$ which
explains why Table 2 has only two columns.

%As already mentioned in Sec.II
The structure functions
combinations $(U^{-}+F^{P-}\pm (U^{P-}+F^{-}))$
are of particular interest since they are directly proportional
to the nonperturbative spin parameter $\epsilon_b$. They
receive contributions from the
helicity $\pm 3/2$ configurations, which are neither populated
by the parton model contribution nor by the spin neutral
kinetic energy term $K_b$. In fact, for the rate combinations
$(U^- + F^{P-})$ and $(U^{P-}+F^-)$ one finds
\bea
 \hat \Gamma_U^{-} +\hat \Gamma_F^{P-}&=&
 -\epsilon_b \hat \Gamma_U^{-}(K_b=0)
\nonumber \\
 &=& -\epsilon_b \left[
  \frac{\sqrt{R}}{3} 
      (1-7  \eta -7  \eta^2+ \eta^3-7  \rho+12  \eta  \rho -7  
         \eta^2  \rho  -7  \rho^2 -7  \eta  \rho^2 +  \rho^3) \right.
 \label{eq:ana1}
 \\ &&
         \left. 
         + 8 \left[ \rho^2 ( \eta^2-1) 
         \ln\left(\frac{1- \eta+ \rho- \sqrt{R}}{2  \sqrt{\rho}}\right)
         + \eta^2 ( \rho^2-1)  
        \ln\left(\frac{1+ \eta- \rho- \sqrt{R}}{2  \sqrt{\eta}}\right) 
                  \right]
    \right]
 \nonumber
\eea
and
\bea
 \hat \Gamma_U^{P-} +\hat \Gamma_F^{-}&=&
 -\epsilon_b \hat \Gamma_F^{-}(K_b=0)
\nonumber \\
 &=& -\epsilon_b \left[
 ( -1+ \eta+2  \sqrt{ \rho} - \rho) ( 1-7  \eta-7  \eta^2+ 
         \eta^3+2  \sqrt{ \rho}-12  \eta  \sqrt{ \rho}
         -2  \eta^2  \sqrt{ \rho} -17  \rho \right.
\nonumber \\ &&
         +38  \eta  \rho  
         -7  \eta^2  \rho +28  \sqrt{ \rho}^3 
         -12  \eta  \sqrt{ \rho}^3-17  \rho^2 -7  \eta  \rho^2+2  
          \sqrt{ \rho}^5+ \rho^3 )/3 
\nonumber \\ && \left.
         +4  \eta^2 (1- \rho)^2 \ln
      \left(\frac{\eta}{(1- \sqrt{ \rho})^2}\right)  
  \right]
 \label{eq:ana2}
\eea
where the analytical results on the r.h.s. of Eqs. (\ref{eq:ana1})
and (\ref{eq:ana2}) can be obtained from the closed form
expressions for $\hat \Gamma_U^{-}$ and $\hat \Gamma_F^{-}$
 in Appendix B by setting $K_b=0$.

Numerically one has
\bea
\hat \Gamma_U^{-}+\hat \Gamma_F^{P-}&=& - 0.0376 \cdot \epsilon_b 
\nonumber \\ 
\hat \Gamma_U^{P-}+\hat \Gamma_F^{-}&=& 0.0208 \cdot \epsilon_b 
\eea

It is clear that the contribution of $K_b$ could be 
extracted from taking appropriate linear combinations
of the fully integrated structure functions listed in Appendix
B. Since the coefficients needed in this extraction involve high 
powers of the masses in the process, which are uncertain,
it would be desirable to find an observable directly proportional
to $K_b$. Such an observable appears in the measurement
of the transverse polarization of the $\tau$ as discussed
in the next section.

\section{Transverse polarization of the $\tau$-lepton}

In this section we present results on the transverse 
polarization components of the $\tau$-lepton in the $(\tau, \bar \nu_\tau)$
c.m. frame.
The two transverse components are conventionally divided into the 
transverse perpendicular component in the lepton-hadron plane
and the transverse normal component out of the lepton-hadron plane.
The latter component is not affected by the boost from
the $\Lambda_b$ rest frame to the  $(\tau, \bar \nu_\tau)$ c.m. frame and,
after the appropriate integrations, can therefore be compared
with the corresponding calculation done in the 
$\Lambda_b$ rest frame in \cite{gremm}.

The $2\times 2$ density matrix of the $\tau$-lepton can be obtained
from the master formula
\bea
 \frac{\mbox{d}\Gamma_{\lambda_l \lambda_l'}}
{\mbox{d}q_0 \mbox{d}q^2 \mbox{d}\cos\Theta \mbox{d}\cos 
\Theta_P \mbox{d}\phi} 
&=&
\frac{2 G^2 \vert V_{bj}\vert^2 (q^2-m_\tau ^2)\sqrt{q_0^2-q^2}}{3 (2\pi)^4
 q^2} \{W(\Theta, \phi, \Theta_P)\}_{\lambda_l \lambda_l '} 
 \mbox{ , }
\label{eq:c1}
\eea
where \cite{bialas}
\bea
\{W(\Theta, \phi, \Theta_P)\}_{\lambda_l \lambda_l '}
&=&\frac{3}{16}\sum_{m,m',J,J',\lambda_b, \lambda_b '}
 \left[ (-1)^{J+J'}
 h_{\lambda_l \frac{1}{2}}
 h^\ast _{\lambda_l ' \frac{1}{2}} e^{-i (m-m')\phi } \right.
\nonumber \\
&& \left. \cdot
 d_{m,\lambda_l-\frac{1}{2}}^{J}(\pi-\Theta)
 d_{m',\lambda_l ' -\frac{1}{2}}^{J'}(\pi-\Theta)
 W^{\lambda_b \lambda_b '}_{m m'} \rho_{\lambda_b \lambda_b '}(\Theta_P)
 \right] \mbox{ . }
\label{eq:c2}
\eea
The helicity amplitudes $h_{\lambda_l \frac{1}{2}}$
($\lambda_l =\pm \frac{1}{2}$) for the decay 
$W^-_{off-shell}\to \tau^-  + \bar \nu_\tau $
appearing in the master formula are given by \cite{koernerschuler}
\bea
 h_{-\frac{1}{2} \frac{1}{2}}&=& \sqrt{8 (q^2-m_\tau ^2)}
= 2 \sqrt{2} m_b \sqrt{\hat q^2-\eta} 
\nonumber \\
 h_{\frac{1}{2} \frac{1}{2}}&=& \sqrt{\frac{m_\tau ^2}{2 q^2}} 
    h_{-\frac{1}{2} \frac{1}{2}}
= \sqrt{\frac{\eta}{2 \hat q^2}} 
    h_{-\frac{1}{2} \frac{1}{2}} 
\mbox{ . }
\label{eq:c3}
 \eea
The diagonal elements of the density matrix
$\{ W \}_{\lambda_l \lambda_l '}$
have already been written down
in the main text. Here we list the nondiagonal density matrix element
relevant for the transverse polarization
components of the $\tau $.

For the unnormalized transverse perpendicular component of the
polarization vector one needs to calculate
$\{ W \}^x = \{ W \}_{+-}+\{ W \}_{-+}$. One has
\bea
\{ W \}^x &=& \{ W(\Theta, \phi, \Theta_P) \}_{+-}+
\{ W(\Theta, \phi, \Theta_P) \}_{-+} =
 \frac{m_\tau }{\sqrt{2 q^2}} (q^2-m_l^2)
\nonumber \\
&& \left[ \frac{3}{2 \sqrt{2}}
 (\rho_{++}(W_{++}^{++}-W_{--}^{++})+\rho_{--}(W_{++}^{--}-W_{--}^{--}))
 \sin \Theta \right.
\nonumber \\
&& -\frac{3}{4 \sqrt{2}}
 (\rho_{++}(W_{++}^{++}+W_{--}^{++}-2 W_{00}^{++})
 +\rho_{--}(W_{++}^{--}+W_{--}^{--}-2 W_{00}^{--}))
 \sin 2\Theta
 \nonumber \\
 && -\frac{3}{\sqrt{2}}
 (\rho_{++} W_{0t}^{++}+ \rho_{--} W_{0t}^{--}) \sin \Theta
 \nonumber \\
 && -\frac{3}{2}\rho_{+-} (W_{0+}^{+-}-W_{-0}^{+-}) \cos 2 \Theta\cos\phi
 \nonumber \\
 && +\frac{3}{2} \rho_{+-} (W_{0+}^{+-}+W_{-0}^{+-}
 +W_{t+}^{+-}-W_{-t}^{+-}) \cos \Theta \cos \phi
 \nonumber \\
 && \left. -\frac{3}{2} \rho_{+-}(W_{t+}^{+-}+W_{-t}^{+-})\cos\phi\right]
\label{eq:c4}
\eea
Note that the $14^{th}$ structure function combination
$(W_{t+}^{+-}+W_{-t}^{+-})$
that was missing from the rate expressions in the main text makes its first
appearance in the transverse normal polarization.
As Eq.(\ref{eq:c4}) shows, $\Lambda_b$-polarization as well as a
determination of the $\tau$-lepton's transverse polarization
is necessary for a determination of the complete set of
14 structure functions.

The unnormalized transverse normal component is given by the
combination $i(W_{+-}-W_{-+})$. One obtains
\bea
\{W\}^y &=& i(\{ W(\Theta, \phi, \Theta_P) \}_{+-}-
\{ W(\Theta, \phi, \Theta_P) \}_{-+} )
\nonumber \\
&=&
 \frac{m_\tau }{\sqrt{2 q^2}} (q^2-m_\tau ^2)
\left[ -\frac{3}{2}
\rho_{+-} (W_{0+}^{+-}+W_{-0}^{+-}+W_{t+}^{+-}-W_{-t}^{+-}) \sin \phi \right.
\nonumber \\
&& \left. +\frac{3}{2} \rho_{+-} (W_{0+}^{+-}-W_{-0}^{+-}
 +W_{t+}^{+-}+W_{-t}^{+-}) \cos \Theta \sin \phi
 \right]
\label{eq:c5}
\eea
Note again the contribution from the
$14^{th}$ structure function combination
$(W_{t+}^{+-}+W_{-t}^{+-})$. It is quite remarkable that the two
combinations of helicity structure functions appearing in the
transverse normal polarization of the $\tau$ in Eq. (\ref{eq:c5})
can be seen to be
entirely determined by the nonperturbative $K_b$ contribution.
We have no simple explanation of this fact.

The $q_0$-integration of the relevant density matrix elements
 (\ref{eq:c1}) can easily be done as described in the main text.
In fact, the relevant $q_0$-integration of the
$14^{th}$ structure function combination is listed in
Appendix A.
The remaining $q^2$-integration is then done numerically.
The numerical result is presented for the case $b\rightarrow c$
using the same numerical values as in Sec.III.
We obtain
\bea
 \frac{\mbox{d}\Gamma^x}{\mbox{d}\cos\Theta \mbox{d}\cos \Theta_P 
\mbox{d}\phi}
 &=& \frac{\Gamma_b}{4 \pi}
 \left[ -0.03777 \sin \Theta + 0.00725 \sin 2\Theta \right.
 \nonumber \\
 && + P \cos \Theta_P (0.01519 \sin \Theta +0.01907 \sin 2\Theta )
 \nonumber \\
 && \left.  + P \sin \Theta_P \cos \phi(0.01682 +0.00091 \cos\Theta
                   +0.01714 \cos 2\Theta )\right]
\label{eq:c6}\\
 \frac{\mbox{d}\Gamma^y}{\mbox{d}\cos\Theta \mbox{d}\cos \Theta_P 
\mbox{d}\phi}
 &=& \frac{\Gamma_b}{4 \pi}
 (-0.000908+0.000323 \cos \Theta ) P \sin \Theta_P \sin \phi
\label{eq:c7}
\eea
The transverse perpendicular polarization is dominated by the
zeroth order parton contribution with angular coefficients
comparable to the entries in Tables 1 and 2. The transverse
normal angular coefficients are quite small as expected since
they are proportional to $K_b$.

As mentioned before the transverse normal polarization is not
affected by the boost from the lab frame to the
$(\tau^- \bar \nu_\tau )$ c.m. frame. In order to compare our results
with the corresponding results in Ref.\cite{gremm} we
integrate the transverse normal polarization with respect to
$\cos \Theta$ and $\cos \Theta_P$ to obtain
\bea
  \frac{\mbox{d}\Gamma^y}{\mbox{d} \hat q^2 \mbox{d}\phi}
 &=& -\frac{\Gamma_b}{2 \pi}
 12\pi P
 \frac{(\hat q^2-\eta)^2}{\hat q^2}
 \sqrt{\frac{\eta}{2\hat q^2}}
 I(W_{0+}^{+-}+W_{-0}^{+-}+W_{t+}^{+-}-W_{-t}^{+-})\sin \phi
\label{eq:c8}
\eea
As noted above, the linear combination
$I(W_{0+}^{+-}+W_{-0}^{+-}+W_{t+}^{+-}-W_{-t}^{+-})$ is proportional
to $K_b$ only and has no zeroth order partonic contribution.
The $\hat q^2$-integration is easily done and one finds
\bea
  \frac{\mbox{d}\Gamma^y}{\mbox{d}\phi}
 &=& \frac{\Gamma_b}{2 \pi} P A \sin\phi
\label{eq:c9}
\eea
where
\bea
 A&=& 2 \pi K_b \sqrt{\eta} \left[
\vphantom{\ln\left(\frac{1- \eta+ \rho- \sqrt{R}}{2  \sqrt{\rho}}\right)
         }
       \sqrt{R}(-2-5\eta +\eta^2 -5\rho+\rho^2+10\eta \rho )/3
       \right.
\nonumber\\
 &&
-4 \eta(1-2\rho+\eta \rho+\rho^2)
 \ln\left(\frac{1+ \eta- \rho- \sqrt{R}}{2  \sqrt{\eta}}\right)
\nonumber\\
 &&
 \left.
 -4 \rho(1-2\eta+\eta\rho+\eta^2)
 \ln\left(\frac{1- \eta+ \rho- \sqrt{R}}{2  \sqrt{\rho}}\right)
 \right]
\label{eq:c10}
\eea
This result agrees with the result in \cite{gremm} when their
 corresponding $y$-distribution is integrated with respect to $y$.
Numerically one has
\bea
A = -0.11 K_b
\eea

The results on the $\tau$ polarization
discussed in the last two sections refer to the $(\tau, \bar \nu_\tau)$
rest frame which, because of the elusiveness of the
neutrino $\bar \nu_\tau$, may not always be easy to construct.
For some applications it may be preferable to avail of the longitudinal
polarization of the $\tau$ in the $\Lambda_b$ rest frame. This is 
the subject of the next section.

\section{Polarization correlations in the $\Lambda_b$ rest frame}

In this section we 
determine the polarization of the $\tau$ in the $\Lambda_b$ rest frame
using a calculational technique originally 
proposed in \cite{ShVo} which involves an averaging over the
azimuthal angle $\phi$ (see Fig.1). 
In this way we determine the polarization
components of the $\tau$ in the plane spanned by the
$\tau $ and the polarization vector of the $\Lambda_b$.

The method of \cite{ShVo} is based on applying the unitarity relation to the
amplitude for forward $\bar\tau \Lambda_b\to \bar\tau \Lambda_b$
scattering 
(see Fig.2)
\bea\label{a1}
T(s) = i\langle \Lambda_b(v,s)\bar\tau(p_\tau)|\int\mbox{d}x
\mbox{T}{\cal H}_W^\dagger (x){\cal H}_W(0)|\Lambda_b(v,s)\bar\tau(p_\tau)
\rangle\,,
\eea
where 
\bea\label{a2}
{\cal H}_W = 2\sqrt{2} V_{jb}G_F\, [\bar j\gamma_\mu P_L b]
[\bar e\gamma^\mu P_L\nu_e]\qquad (P_L=\frac12(1-\gamma_5))
\eea
is the interaction Hamiltonian responsible for the semileptonic decay
$b\to j \tau^- \bar\nu_\tau$ and $s=(m_{\Lambda_b}v+p_\tau)^2$. By inserting a 
complete set of states in (\ref{a1}) one obtains
\bea\label{a3}
\mbox{Im }T(s) &=& \frac12\sum_{X_j,\nu_\tau}\int\mbox{d}\mu(X_j)\mbox{d}
\mu(\nu_\tau) (2\pi)^4 \delta^4(m_{\Lambda_b}v+p_\tau-p_{X_j}-p_{\nu_\tau})\\
& &\times\,|\langle X_j\bar\nu_\tau|
{\cal H}_W(0)|\Lambda_b(v,s)\bar\tau(p_\tau)\rangle|^2\,.\nonumber
\eea
The phase-space volume element is $\mbox{d}\mu = \frac{d^3p}{(2\pi)^32E}$.

Comparing (\ref{a3}) with the inclusive semileptonic decay rate
\bea
\mbox{d}\Gamma = |\langle X_j\bar\nu_\tau(p_{\nu_\tau})\tau(p_\tau)|
{\cal H}_W|\Lambda_b(v,s)\rangle|^2 (2\pi)^4\delta^4(m_{\Lambda_b}v-
p_\tau-p_{X_j}-p_{\nu_\tau})\mbox{d}\mu(X_j)\mbox{d}\mu(\nu_\tau)\mbox{d}
\mu(\tau)
\eea
one obtains the final formula relating (\ref{a1}) to quantities of 
experimental interest:
\bea\label{a5}
\mbox{d}\Gamma = \frac{1}{(2\pi)^3} |\vec p_\tau |\mbox{d}E_\tau
\mbox{d}\Omega_\tau\,\mbox{Im }T(s)\,.
\eea

Our problem is thus reduced to computing the function Im $T(s)$. This can
be done with the help of an operator-product expansion combined with 
the heavy mass expansion as discussed in Sec.II. 
First the heavy
hadron is replaced with a heavy quark with momentum $p_b=m_bv+k$.
From the 1-loop diagram in Fig.3 one reads off the lowest-order expression
for $T(s)$
\bea\label{a6}
T(s) &=&  8G_F^2|V_{jb}|^2
\left[ \bar u(p_\tau)\gamma_\mu \gamma^\alpha \gamma_\nu P_Lu(p_\tau)
\right]
\left[\bar u(v)\gamma^\nu \gamma^\beta \gamma^\mu P_L u(v)\right]
I_{\alpha\beta}(k)
\eea
where we have defined
\bea\label{a7}
I_{\alpha\beta}(k) = \int\frac{\mbox{d}^nq}{(2\pi)^n}
\frac{(p_\tau-q)_\alpha (m_bv+k-q)_\beta}
{[(p_\tau-q)^2+i\epsilon][(m_bv+k-q)^2-m_c^2+i\epsilon]}\,.
\eea

We will be interested in polarized $\tau$ leptons in the final state.
Thus the leptonic amplitude in (\ref{a6}) should be replaced by
\bea\label{a8}
\bar u(p_\tau)\gamma_\mu \gamma^\alpha \gamma_\nu P_Lu(p_\tau)\to
\mbox{Tr }\left\{\frac{1}{2m_\tau}(\pslash_\tau+m_\tau)\frac12
(1+\gamma_5\sslash_\tau)\gamma_\mu \gamma^\alpha \gamma_\nu P_L\right\}\,,
\eea
with $s_\tau$ the spin vector of the $\tau$ lepton.

The integral $I_{\alpha\beta}(k)$ in (\ref{a7}) can be calculated by 
combining the denominators with a Feynman parameter $x$. The
result is a complex function  with an imaginary part 
given by 
\bea\label{a9}
\frac{1}{2\pi}\mbox{Im }I_{\alpha\beta}(k) &=&\\
& &\hspace*{-2cm}\frac{1}{(4\pi)^2}\int_0^1\mbox{d}x
\left\{ \frac12 g_{\alpha\beta}s
- x(1-x)[m_bv+k-p_\tau]_\alpha [m_bv+k-p_\tau]_\beta\right\}\theta(x_1-x)
\nonumber
\eea
where
\bea\label{a10}
s = x\left\{ -m_\tau^2(1-x) + m_c^2 - (m_bv+k)^2(1-x) +
   2(1-x)p_\tau\cdot(m_bv+k)\right\}\,.
\eea
In (\ref{a9}) we have denoted $x_1$ the root of the equation $s(x_1)=0$.
It is given by 
\bea\label{a11}
x_1 &=& 1 - \frac{\displaystyle\rho}{\displaystyle (1+\eta-y)\left(1 +
\frac{\displaystyle2\tilde k\cdot
(v-\tilde p_\tau) + \tilde k^2}{\displaystyle 1+\eta-y}\right)}\\
 &=&
x_0 + \frac{1-x_0}{1+\eta-y}\left(2\tilde k\cdot(v-\tilde p_\tau) +
\tilde k^2 - \frac{4[\tilde k\cdot(v-\tilde p_\tau)]^2}{1+\eta-y} +\cdots
\right)\nonumber\,,
\eea
where $x_0=1-\rho/(1+\eta-y)$ is the value of $x_1$ for $k=0$. We have 
introduced reduced momenta $\tilde k=k/m_b$, $\tilde p_\tau=p_\tau/m_b$
and have expanded in powers of $\tilde k$ up to second order.

The integration in (\ref{a9}) can easily be performed with the result
\bea
\mbox{Im }I_{\alpha\beta}(k) &=& \frac{1}{8\pi}\left\{\frac14\rho x_1^2
+ \frac12 \left(\frac{x_1^2}{2}-\frac{x_1^3}{3}\right)\left[
-\eta-(v+\tilde k)^2+2\tilde p_\tau\cdot (v+\tilde k)\right]\right\}
g_{\alpha\beta}\nonumber\\\label{a12}
&-& \frac{1}{8\pi}\left(\frac{x_1^2}{2}-\frac{x_1^3}{3}\right)
(v+\tilde k-\tilde p_\tau)_\alpha (v+\tilde k-\tilde p_\tau)_\beta\,,
\eea
where $x_1$ has to be replaced with 
the expanded form of Eq.(\ref{a11}).

In physical applications we are interested in heavy hadron decay, rather 
than free heavy quark decay. One can obtain the corresponding scattering
amplitude $T(s)$ by replacing the hadronic spinor expression in (\ref{a6})
with expectation values of the appropriate operators
\bea\label{a13}
\bar u(v) f(\tilde k_\alpha)u(v) \to \langle\Lambda_b(v,s)|
\bar bf\left(\frac{iD_\alpha}{m_b}\right)b|\Lambda_b(v,s)\rangle\,.
\eea
These matrix elements can in turn be expanded in powers of $1/m_b$ with
the help of heavy quark effective theory (HQET) methods as discussed earlier.
Referring to \cite{shifman,man}
  for calculational details,
we only give the final substitution rules needed for computing Im
$T(s)$.

a) terms of order $\tilde k^0$
\bea\label{a14}
& &\gamma_\mu \to v_\mu\\
& &\gamma_\mu\gamma_5 \to s_\mu(1+\epsilon_b)
\eea

b) terms of order $\tilde k^1$
\bea
& &\tilde k_\mu \gamma_\nu \to \frac13 K_b(-2g_{\mu\nu}+5v_\mu v_\nu)\\
& &\tilde k_\mu \gamma_\nu\gamma_5 \to K_b (v_\mu s_\nu + \frac23
v_\nu s_\mu)
\eea

c) terms of order $\tilde k^2$
\bea
& &\tilde k_\mu\tilde k_\nu \gamma_\alpha \to
 -\frac23 K_b(g_{\mu\nu}-v_\mu v_\nu)v_\alpha\\
& &\tilde k_\mu\tilde k_\nu \gamma_\alpha\gamma_5 \to
-\frac23 K_b (g_{\mu\nu}-v_\mu v_\nu)s_\alpha\label{a19}
\eea

It is now a simple matter to combine (\ref{a11},\ref{a12}) and 
(\ref{a14}-\ref{a19}) into (\ref{a5}) and extract the decay rate.
It can conveniently be written by splitting the rate into two terms
\bea\label{a20}
\mbox{d}\Gamma = \mbox{d}\Gamma + \mbox{d}\Gamma^p =
\frac12\left(\mbox{d}\Gamma(s_\tau)+\mbox{d}\Gamma(-s_\tau)\right) +
\frac12\left(\mbox{d}\Gamma(s_\tau)-\mbox{d}\Gamma(-s_\tau)\right)\,.
\eea
d$\Gamma$ and 
d$\Gamma^p$ represent the decay rates into unpolarized
and polarized leptons. For the unpolarized rate we obtain
\bea
\frac{1}{\Gamma_b}\frac{\mbox{d}\Gamma}{\mbox{d}y\mbox{d}\cos\theta_\tau}
&=&\sqrt{y^2-4\eta} \left( A(y) + (\tilde p_\tau \cdot s) B(y) \right)
\,,
\label{eq:separat}
\eea
where 
\bea
A(y)&=&x_0^2(-3y^2+6y(1+\eta)-12\eta) + x_0^3(y^2-3y(1+\eta)+8\eta)\\
&+&\left. K_b\left\{ 2[(1+\eta)^2-y^2] - 4x_0[y^2 - y(1+\eta) + 2(1+\eta^2)]
\right.\right.\nonumber\\
& &\hspace{-2cm}+\left.\left.x_0^2[4y^2-8y(1+\eta)+16\eta+10(1+\eta^2)] +
x_0^3[-\frac43 y^2+4y(1+\eta)-\frac{32}{3}\eta-4(1+\eta^2)]\right.\right.
\nonumber\\
&&
\hspace{-2cm}
-\left.\left.4\frac{(1-\eta)^2(1+\eta)(1-x_0)^2(1-3x_0)}
{1+\eta-y} +
2\frac{(1-\eta)^4(1-x_0)^2(1-4x_0)}{(1+\eta-y)^2}\right\}\right.\nonumber\\
B(y)&=&
\left. 
%\frac12\sqrt{y^2-4\eta}\cos\theta
\left\{(1+\epsilon_b)
[6x_0^2(y-2\eta)-2x_0^3(1+y-3\eta)\right.\right.\nonumber\\
& &\hspace{-2cm}\left.\left.+K_b\left[4(y+2)-8x_0(2+\eta-y) + 
8x_0^2(1+2\eta-y)+\frac83 x_0^3(y-3\eta)
\right.\right.\right.\nonumber\\
& &\hspace{-2cm}\left.\left.- 4\frac{(1-\eta)(1-x_0)^2[3+\eta-2x_0
(\eta+2)]}
{1+\eta-y} + 4\frac{(1-\eta)^3(1-x_0)^2(1-4x_0)}{(1+\eta-y)^2}
\right]\right\}\,.\nonumber
\eea
 We have defined $\theta_\tau$ as the angle between the $\Lambda_b$ spin
and the $\tau$ lepton momentum direction.
 For the decay rate into polarized leptons d$\Gamma^p$
we obtain 
\bea
2\frac{1}{\Gamma_b}\frac{\mbox{d}\Gamma^p}{\mbox{d}y\mbox{d}\cos\theta_\tau}
= \sqrt{y^2-4\eta}\,
\sqrt{\eta} \left\{(v\cdot s_\tau)(\tilde p_\tau\cdot s) A^p(y)
+ (v\cdot s_\tau) B^p(y) + (s\cdot s_\tau) C^p(y)\right\}\,,
\eea
where 
\bea
A^p(y) &=& -24(1+\epsilon_b)
\left(\frac{x_0^2}{2}-\frac{x_0^3}{3}\right)
- 4K_b\left[ 2+2x_0-5x_0^2+\frac83 x_0^3\right.\\\nonumber
& &\left. - 2\frac{(1+\eta)(1-x_0)^2(2-x_0)}
{1+\eta-y} + 2\frac{(1-\eta)^2(1-x_0)^2(1-4x_0)}{(1+\eta-y)^2}\right]\\
B^p(y) &=& -6x_0^2(2-y) + 2x_0^3(3-y-\eta) - 4K_b\left[ 
-y-2\eta+2x_0(1+2\eta-y)\right.\\\nonumber
&+&\left. 2x_0^2(-2-\eta+y) + \frac23 x_0^3(3-y) +
\frac{(1-\eta)(1-x_0)^2[-1-3\eta+2x_0(1+2\eta)]}{1+\eta-y}\right.\\\nonumber
& &\left.+\,
\frac{(1-\eta)^3(1-x_0)^2(1-4x_0)}{(1+\eta-y)^2}\right]\nonumber\\
C^p(y) &=& 2x_0^3(1+\epsilon_b)(1+\eta-y) - 2K_b\left[
2x_0(1+\eta+y) + x_0^2(-4-4\eta+y)\right.\\\nonumber
& &\left. + x_0^3(2+2\eta-\frac43y) - \frac{2x_0(1-\eta)^2(1-x_0)^2}
{1+\eta-y}\right]\,.
\eea

We have checked that this formula agrees with the various particular cases
presented in the literature. Thus, \cite{FaLi} compute the longitudinal
polarization asymmetry corresponding to an unpolarized decaying baryon
and \cite{gremm} give the full results for various lepton polarizations
from the decay of 
polarized $\Lambda_b$ baryons. The method presented here has the
advantage of performing the integration over the neutrino phase space 
automatically. Unfortunately, this aspect can prove to
be also a limitation: because of integrating over all possible neutrino
momenta, all information about the decay plane is lost. Therefore this
method can only be applied for obtaining lepton polarization asymmetries
averaged over the  position of the decay plane, i.e. after azimuthal 
averaging. The azimuthal averaging does not affect the longitudinal
polarization component and thus our results can directly be compared
to corresponding results in the literature
\cite{gremm,FaLi}.
We have also checked that  we agree with \cite{gremm} on the transverse
polarization component in the plane spanned by the
$\vec p_\tau$ and $\vec P$ which can be obtained from \cite{gremm}
with the appropriate azimuthal averaging. The transverse polarization
component normal to this plane averages out when doing the 
azimuthal averaging.
Our results for the transverse normal component of the 
$\tau$-polarization have been presented in Sec.IV.
 
\section{Summary and Conclusions} 

We have analyzed the inclusive semileptonic decays of polarized
$\Lambda_b$ baryons into polarized $\tau$-leptons.
We discussed spin-spin, spin-momentum and momentum-momentum correlations
between the spins of the $\Lambda_b$ and the $\tau$, and the momenta
of the virtual $W$ (or recoil momenta $\vec p_X$) and the $\tau$.
Using helicity techniques we presented detailed results on the
above angular correlations involving a three-fold angular decay
distribution in the three helicity angles that can be defined for the
process. By taking suitable combinations of helicity structure functions
we identified observables that are directly proportional to the contributions
of the $O(1/m_b^2)$ nonperturbative matrix elements. In the helicity
method one determines the $\tau$-polarization in 
the ($\tau^-,\nu_\tau$) rest frame.
We give also results on the $\tau$-polarization in the
$\Lambda_b$-rest frame using an elegant loop calculation.

\section*{Acknowledgment}

We would like to thank M. Neubert for an instructive discussion. 
S. B. thanks H. Druxes for generously providing access to 
computer facilities at the University of Koblenz. The research of D.P.
was supported by the Ministry of Science and the Arts of Israel.
\newpage

\section*{Appendix A: $q_0$-Integrated Structure Functions} 

In this appendix we list the results of integrating the structure functions  
in Eq.(\ref{eq:helamplituden}) with respect to $q_0$. 
Our results are given in terms of the integrals 
\bea 
 m_b \int \sqrt{\hat q_0^2-\hat q^2} W d\hat q_0 &=& I(W) 
 \nonumber  
\eea 
where $W$ stands for any of the linear combinations of the helicity  
structure functions in Eq.(\ref{eq:helamplituden}). One has 
\bea 
%1 
I( W_{++}^{++} + W_{--}^{++} + W_{++}^{--} + W_{--}^{--})  
&=& 
(1- K_b) \hat p ( - \hat q^2 + \rho + 1)/2 
+4 K_b \hat p /3 
\nonumber \\ 
&&+G_b( \hat p^2(15(\hat q^2 - \rho) - 11)+ \hat q^2
(3 \hat q^2
 - 3 \rho) - 7) - 4 \rho + 4))/(6 \hat p)
\nonumber \\ 
%2 
I( W_{++}^{++} + W_{--}^{++} - W_{++}^{--} - W_{--}^{--})  
&=& 
(1+\epsilon_b) \hat p^2 
+2 K_b ( - 2 \hat p^2 - q^2)/3) 
\nonumber \\ 
%3 
I(W_{00}^{++} + W_{00}^{--}) 
&=& 
(1- K_b) \hat p (4 \hat p^2 - \hat q^4 + \hat q^2 \rho + \hat q^2)/(4 \hat q^2) 
- 4 K_b \hat p /3
\nonumber \\ 
&&+G_b\left(\hat p^2(15(-4 \hat p^2 +\hat q^4 -\hat q^2 \rho )
 -59 \hat q^2 - 12 \rho + 12)
 \right.
\nonumber \\
&&
 \left. \mbox{\hspace*{.6cm}}
 + \hat q^2 (\hat q^2 (3(\hat q^2 -\rho) - 7) - 4 \rho + 4)\right)
 /(12 \hat p \hat q^2)
\nonumber \\ 
%4 
I(W_{00}^{++} - W_{00}^{--}) 
&=& 
(1+\epsilon_b) \hat p^2 (\rho - 1)/(2 \hat q^2) 
+K_b ( - 2 \hat p^2 - \rho + 1)/3 
\nonumber \\ 
%5 
I(W_{++}^{++}- W_{--}^{++} + W_{++}^{--}- W_{--}^{--})  
&=& 
-\hat p^2  
+2 K_b (2 \hat p^2 + \hat q^2)/3 
+ G_b (10 \hat p^2 + 2 \hat q^2 + 3 \rho - 3)/3
\nonumber \\ 
%6 
I(W_{++}^{++}- W_{--}^{++} - W_{++}^{--}+ W_{--}^{--})  
&=& 
(1+\epsilon_b) \hat p (\hat q^2 - \rho - 1)/2 
+K_b \hat p ( - 3 \hat q^2 + 3 \rho - 5)/6 
\nonumber \\ 
%7 
I(W_{-0}^{+-} + W_{0+}^{+-}) 
&=& 
(1+\epsilon_b) \hat p (\hat q^2 + \rho - 1)/(2 \sqrt{2}\sqrt{\hat q^2}) 
\nonumber \\
&&
+K_b \hat p (\hat q^2 - \rho + 1)/(3 \sqrt{2}\sqrt{\hat q^2}) 
\nonumber \\ 
%8 
I(W_{-0}^{+-} - W_{0+}^{+-}) 
&=& 
-(1+\epsilon_b) \hat p^2/( \sqrt{2}\sqrt{\hat q^2}) 
+K_b ( - \hat p^2 + 2 \hat q^2)/(3 \sqrt{2}\sqrt{\hat q^2}) 
\nonumber \\ 
%13 
I(W_{tt}^{++} + W_{tt}^{--}) 
&=& 
(1-K_b) \hat p (4 \hat p^2 - \hat q^4 +  
\hat q^2 \rho + \hat q^2)/ (4 \hat q^2) 
\nonumber \\ 
&&
 +G_b \left(\hat p^2(5(-4\hat p^2 + \hat q^4 - 5 \hat q^2 \rho)
  - 9 \hat q^2 - 4 \rho + 4 ) \right.
\nonumber \\ 
&& \left. \mbox{\hspace*{.6cm}}
+ \hat q^4(\hat q^2 -\rho - 1)\right)/(4 \hat p \hat q^2)
\nonumber \\
%14 
I(W_{tt}^{++} - W_{tt}^{--}) 
&=& 
(1+\epsilon_b) \hat p^2 (\rho - 1)/(2 \hat q^2) 
\nonumber \\ 
&&  
+K_b (  - 2 \hat p^2 \rho + 2 \hat p^2 - \hat q^2 \rho +  
\hat q^2)/(3 \hat q^2) 
\nonumber \\ 
%15 
I(W_{0t}^{++} + W_{0t}^{--}) 
&=& 
\hat p^2 ( - \rho + 1)/(2 \hat q^2) 
+ K_b (2 \hat p^2 \rho - 2 \hat p^2 +  
\hat q^2 \rho - \hat q^2)/(3 \hat q^2) 
\nonumber \\
&& + G_b
(10 \hat p^2 \rho - 2 \hat p^2 + 5 \hat q^2 \rho -
\hat q^2)/(6 \hat q^2)
\nonumber \\
%16 
I(W_{0t}^{++} - W_{0t}^{--}) 
&=& 
(1+\epsilon_b) \hat p ( - 4 \hat p^2 + \hat q^4 -  
\hat q^2 \rho - \hat q^2)/(4 \hat q^2) 
\nonumber \\ 
&&  
+K_b \hat p (12 \hat p^2 - 3 \hat q^4 +  
3 \hat q^2 \rho + 11 \hat q^2)/(12 \hat q^2) 
\nonumber \\ 
%17 
I(W_{-t}^{+-} - W_{t+}^{+-}) 
&=& 
(1+\epsilon_b) \hat p (\hat q^2 + \rho - 1)/(2 \sqrt{2}\sqrt{\hat q^2}) 
\nonumber \\ 
&& 
+K_b \hat p (-\hat q^2 + \rho - 1)/(3 \sqrt{2}\sqrt{\hat q^2}) 
\nonumber  \\
%19 
I(W_{-t}^{+-} + W_{t+}^{+-}) 
&=& 
(1+\epsilon_b) (-\hat p^2)/(\sqrt{2}\sqrt{\hat q^2}) 
\nonumber \\ 
&& 
+K_b (3\hat p^2 +2\hat q^2 )/(3 \sqrt{2}\sqrt{\hat q^2}) 
\nonumber  
\eea 
where $\hat p=\frac{1}{2}\sqrt{(1-\rho+\hat q^2)^2-4 \hat q^2}$.
For the sake of completeness 
we have retained the contribution of the chromomagnetic interaction 
$G_b$  contribution, although it vanishes for $\Lambda_b$-decays.
For applications of the helicity formalism to the mesonic sector
one has to set $P=0$ in the rate formulae and replace
$( W_{\lambda_W \lambda'_W}^{++} + W_{\lambda_W \lambda'_W}^{--} )$ by
 $W_{\lambda_W \lambda'_W}$.
\newpage

\section*{Appendix B: Fully integrated rate functions} 

In this Appendix we list our fully integrated results for those structure
function components that determine the angular decay distribution for
unpolarized $\Lambda_b$-decay, i.e. for the helicity rates
$\hat\Gamma_U^-$, $\hat\Gamma_L^-$, $\hat\Gamma_F^-$, 
$\hat\Gamma_U^+$, $\hat\Gamma_L^+$,  $\hat\Gamma_S^+$
and  $\hat\Gamma_{SL}^+$.
\bea
\hat\Gamma_U^-&=&
 (1-K_b) \frac{\sqrt{R}}{3} 
      (1-7  \eta -7  \eta^2+ \eta^3-7  \rho+12  \eta  \rho -7  
         \eta^2  \rho  -7  \rho^2 -7  \eta  \rho^2 +  \rho^3)
\nonumber \\ &&
 +16 K_b \frac{\sqrt{R}}{9}
 (1-5 \eta -2 \eta^2 +10 \rho -5 \eta \rho +\rho^2)
\nonumber \\ &&
         + (1-K_b) 8 \left[ \rho^2 ( \eta^2-1) 
         \ln\left(\frac{1- \eta+ \rho- \sqrt{R}}{2  \sqrt{\rho}}\right)
         + \eta^2 ( \rho^2-1)  
        \ln\left(\frac{1+ \eta- \rho- \sqrt{R}}{2  \sqrt{\eta}}\right) 
                  \right]
\nonumber \\ &&
 +K_b\left[
\frac{64 \rho}{3}(1-2 \eta +\eta^2+\rho)
   \ln\left(\frac{1- \eta+ \rho- \sqrt{R}}{2  \sqrt{\rho}} \right)
    \right.
\nonumber \\ && 
     \left.
  + \frac{64 \eta^2}{3} ( \rho-1)   
      \ln\left(\frac{1+ \eta- \rho- \sqrt{R}}{2  \sqrt{\eta}}\right)
    \right]
\\ 
\hat\Gamma_L^-&=&
(1-K_b) \frac{2 \sqrt{R}}{3} (1+10  \eta +  \eta^2-
      7  \rho-10  \eta  \rho +  \eta^2  \rho
               -7  \rho^2 +10  \eta  \rho^2 +  \rho^3)
\nonumber \\ &&
-K_b 
\frac{16\sqrt{R}}{9}(1-5\eta-2\eta^2+10\rho-5\eta \rho+\rho^2)
\nonumber \\ &&
   +(1-K_b) 8  \rho^2 (-2+3  \eta - \eta^2 - \eta  \rho) 
     \ln\left(\frac{1- \eta+ \rho- \sqrt{R}}{2  \sqrt{\rho}}\right)
\nonumber \\ &&
   +K_b \frac{64 \rho}{3}(-1+2 \eta -\eta^2)
 \ln\left(\frac{1- \eta+ \rho- \sqrt{R}}{2  \sqrt{\rho}} \right)
\nonumber \\ &&
         +(1-K_b) 8 \eta (1-\rho)
   (1+\eta-2\rho+\eta\rho+\rho^2)
         \ln\left(\frac{1+ \eta- \rho- \sqrt{R}}{2  \sqrt{\eta}}\right) 
\nonumber \\ &&
   +K_b \frac{64 \eta^2}{3} (1-\rho)
  \ln\left(\frac{1+ \eta- \rho- \sqrt{R}}{2  \sqrt{\eta}}\right) 
\\
\hat\Gamma_F^-&=&
 ( -1+ \eta+2  \sqrt{ \rho} - \rho) ( 1-7  \eta-7  \eta^2+ 
         \eta^3+2  \sqrt{ \rho}-12  \eta  \sqrt{ \rho}
         -2  \eta^2  \sqrt{ \rho} -17  \rho+38  \eta  \rho 
\nonumber \\ &&
         -7  \eta^2  \rho +28  \sqrt{ \rho}^3 
         -12  \eta  \sqrt{ \rho}^3-17  \rho^2 -7  \eta  \rho^2+2  
          \sqrt{ \rho}^5+ \rho^3 )/3 
\nonumber \\ &&
         +4  \eta^2 (1- \rho)^2 \ln
      \left(\frac{\eta}{(1- \sqrt{ \rho})^2}\right)  
\nonumber \\ &&
+K_b\left[
\frac{4}{9} (1 -  \eta - 2  \sqrt{ \rho} +  \rho) 
       (9 - 23  \eta +  \eta^2 +  \eta^3 - 30  \sqrt{ \rho} + 
      20  \eta  \sqrt{ \rho} - 2  \eta^2  \sqrt{ \rho}   \right.
\nonumber \\ &&
 +       31  \rho + 22  \eta \rho
     - 7  \eta^2    \rho 
       - 4  \sqrt{ \rho}^3 - 12  \eta  \sqrt{ \rho}^3 - 
       9  \rho^2 - 7  \eta  \rho^2 + 
      2  \sqrt{ \rho}^5 +  \rho^3)
\nonumber \\ &&
  \left.
   -\frac{16  \eta^2}{3} (1- \rho)^2 
  \ln\left(\frac{\eta}{(1- \sqrt{ \rho})^2}\right) \right]
\\
\hat\Gamma_U^+&=&
(1-K_b) \frac{2 \eta \sqrt{R}}{3} (1+10  \eta +  \eta^2-2  \rho
         +10  \eta  \rho + \rho^2)
+K_b \frac{8 \eta \sqrt{R}}{3} (1+5 \eta+\rho)
\nonumber \\ &&
         +(1-K_b) \frac{8 \eta^2 \rho^2}{1-\rho}
(-1+\eta+\rho)
 \ln\left(\frac{1- \eta+ \rho- \sqrt{R}}{2  \sqrt{\rho}}\right) 
\nonumber \\ &&
+K_b \frac{32 \eta \rho}{3(1- \rho)}
(1-2 \eta+\eta^2-\rho+2\eta \rho)
  \ln\left(\frac{1- \eta+ \rho- \sqrt{R}}{2  \sqrt{\rho}}\right) 
\nonumber \\ &&
  +(1-K_b) \frac{8 \eta^2}{1-\rho}
(1-\rho-\rho^2+\rho^3+\eta+\eta \rho^2)
         \ln\left(\frac{1+ \eta- \rho- \sqrt{R}}{2  \sqrt{\eta}}\right) 
\nonumber \\ &&
+K_b \frac{32 \eta^2}{3(1- \rho)}
 (2+\eta-4 \rho+\eta \rho +2\rho^2)
    \ln\left(\frac{1+ \eta- \rho- \sqrt{R}}{2  \sqrt{\eta}}\right) 
\\
\hat\Gamma_L^+&=&
  (1-K_b) \eta \sqrt{R} (-3-3  \eta +4  \rho -3  \eta  \rho-3  \rho^2)
-K_b \frac{8 \eta \sqrt{R}}{3}
 (1+5\eta+\rho)
\nonumber \\ &&
+(1-K_b)\frac{2\eta \rho^2}{1-\rho}
 (-3+4\eta-\eta^2+4\rho-4\eta\rho-\rho^2)
   \ln\left(\frac{1- \eta+ \rho- \sqrt{R}}{2 \sqrt{\rho}}\right)  
\nonumber \\ &&
+K_b \frac{32 \eta \rho}{3(1- \rho)}
(-1+2 \eta-\eta^2+\rho-2\eta \rho)
  \ln\left(\frac{1- \eta+ \rho- \sqrt{R}}{2  \sqrt{\rho}}\right) 
\nonumber \\ &&
+(1-K_b) 2  \eta (-4  \eta (1- \rho^2)
        +\frac{-1- \eta^2+4  \rho-6  
        \rho^2- \eta^2  \rho^2
         +4  \rho^3- \rho^4}{1- \rho})
      \ln\left(\frac{1+ \eta- \rho- \sqrt{R}}{2  \sqrt{\eta}}\right)
\nonumber \\ &&
+K_b \frac{32 \eta^2}{3(1- \rho)}
 (-2-\eta+4 \rho-\eta \rho -2\rho^2)
    \ln\left(\frac{1+ \eta- \rho- \sqrt{R}}{2  \sqrt{\eta}}\right) 
\\
\hat\Gamma_S^+&=&
   (1-K_b) \left[ \eta  \sqrt{R} 
 (-3-3  \eta +4  \rho -3  \eta  \rho-3  \rho^2)\right.
\nonumber \\ &&
        +2 \eta (-4 \eta (1- \rho^2)+
    \frac{-1- \eta^2+4  \rho-6  
        \rho^2- \eta^2  \rho^2
         +4  \rho^3- \rho^4}{1- \rho})
        \ln\left(\frac{1+ \eta- \rho- \sqrt{R}}{2 \sqrt{\eta}}\right)
\nonumber \\ &&
\left.
      -\frac{2\eta \rho^2}{1 -  \rho} (3 - 4  \eta +  \eta^2 - 4  \rho + 
       4  \eta  \rho +  \rho^2) 
       \ln\left(\frac{1- \eta+ \rho- \sqrt{R}}{2\sqrt{\rho}}\right) \right]
\\
\hat\Gamma_{SL}^+&=&
\eta \frac{1+ \sqrt{ \rho}}{\sqrt{ \rho}-1} (1- \eta-2  
        \sqrt{ \rho}+  \rho) (3+ 3  \eta -2  \sqrt{ \rho}
        -4  \eta  \sqrt{ \rho} -2  \rho +3  \eta  \rho -2  
        \sqrt{ \rho}^3+3  \rho^2)
\nonumber \\ &&
        - \eta (1+4   \eta +  \eta^2 -3  \rho- \eta^2  \rho+3  \rho^2
              -4  \eta  \rho^2- 
        \rho^3) \ln\left(\frac{\eta}{(1- \sqrt{ \rho})^2}\right) 
\nonumber \\ &&
+K_b\left[ 
 \frac{4\eta}{3} (1 +  \sqrt{ \rho}) (-1 +  \eta + 2  \sqrt{ \rho} 
     -  \rho) 
    (-1 -  \eta - 3  \sqrt{ \rho} + 3  \eta  \sqrt{ \rho} 
     +  \rho + 3  \sqrt{ \rho}^3) 
\right.
\nonumber \\ &&
\left.
+\frac{4\eta}{3} (\rho -1 ) (-1 -  \eta^2 + 2  \rho - 4  \eta  \rho 
       -  \rho^2) \ln\left(\frac{\eta}{(1- \sqrt{ \rho})^2}\right)\right]
\eea
where $R$ is related to the maximal momentum of the $\tau$ in the
$\Lambda_b$ rest frame and is given by 
\bea
 R&=&1-2  \eta+ \eta^2-2  \rho-2  \eta  \rho + \rho^2
\nonumber 
\eea
As a necessary check we reproduce the total rate formula
\cite{zal2,zal3,hagiwara,gremm}:
\bea
 \Gamma &=&\Gamma_b \left( \hat\Gamma_U^- + \hat\Gamma_U^+
           +\hat\Gamma_L^- + \hat\Gamma_L^+
           +3 \hat\Gamma_S^+ \right)
\nonumber\\
 &=& \Gamma_b \left(1-K_b\right) \left[
\vphantom{\ln\left(\frac{1- \eta+ \rho- \sqrt{R}}{2  \sqrt{\rho}}\right)
         }
       \sqrt{R}(1-7  \eta -7  \eta^2+ \eta^3-7  \rho+12  \eta  \rho -7  
         \eta^2  \rho  -7  \rho^2 -7  \eta  \rho^2 +  \rho^3)
       \right.
\nonumber\\
 &&
 \left.
   +24 \eta^2(\rho^2-1) 
 \ln\left(\frac{1+ \eta- \rho- \sqrt{R}}{2  \sqrt{\eta}}\right)
   +24 \rho^2(\eta^2-1)
 \ln\left(\frac{1- \eta+ \rho- \sqrt{R}}{2  \sqrt{\rho}}\right)
 \right]
\eea
We mention that we have double checked all our analytic results
by comparing them to a numerical evaluation.

\newpage 

\begin{figure}
\begin{center}
\begin{picture}(0,0)%
\includegraphics{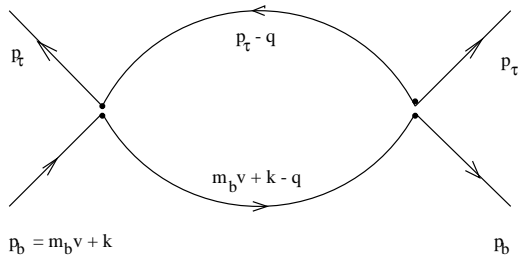}%
\end{picture}%
\setlength{\unitlength}{0.012500in}%
\begin{picture}(213,104)(252,545)
\end{picture}
\end{center}
\caption{Lowest-order graph contributing to the forward scattering 
amplitude $T(s)$.}
\label{fig3}
\end{figure}


\newpage
 

 
\begin{thebibliography}{12} 
\bibitem{close} F.E. Close, J.G. K\"orner, R.J.N. Phillips and
	D.J.Summers, {\em J.Phys.}{\bf G18} (1992) 1716;
    J.G. K\"orner, A. Pilaftsis and M.M. Tung, 
     {\em Z.Phys.}{\bf C63} (1994) 575;
    S. Groote, J.G. K\"orner and M.M. Tung,
         {\em Z.Phys.}{\bf C} (to be published). 
\bibitem{bialas} P. Bialas, J.G. K\"orner, M. Kr\"amer and K. Zalewski,
   {\em Z.Phys.}{\bf C57} (1993) 115. 
\bibitem{kraemer} J.G. K\"orner and M. Kr\"amer, 
   {\em Phys.Lett.}{\bf B275} (1992) 495. 
\bibitem{koernerschuler} J.G. K\"orner and G.A. Schuler, 
   {\em Z.Phys.}{\bf C46} (1990) 93.
\bibitem{zalewski1} A. Kotanski and K. Zalewski, 
   {\em Nucl.Phys.}{\bf B4} (1968) 559. 
\bibitem{zal2} A. Kotanski and K. Zalewski, 
   {\em Nucl.Phys.}{\bf B20} (1970) 236. 
\bibitem{zal3} A. Kotanski and K. Zalewski, 
   {\em Nucl.Phys.}{\bf B22} (1970) 317. 
\bibitem{hagiwara} K. Hagiwara, A.D. Martin and M.F. Wade,
   {\em Phys.Lett.}{\bf B228} (1989) 144;
   {\em Nucl.Phys.}{\bf B327} (1989) 569. 
\bibitem{gremm} M. Gremm, G. K\"opp and L.M. Sehgal, 
   {\em Phys.Rev.}{\bf D52} (1995) 1588. 
\bibitem{man} A.V. Manohar and M.B. Wise, {\em Phys.Rev.}{\bf D49} (1994) 1310.
\bibitem{chay} J. Chay, H.M. Georgi and B. Grinstein, {\em Phys.Lett.}{\bf B247}
   (1990) 399.
\bibitem{bigi} I.I. Bigi, M. Shifman, N.G. Uraltsev and A.I. Vainshtein,
   {\em Phys.Rev.Lett.}{\bf 71} (1993) 496.
\bibitem{shifman} B. Blok, L. Koyrakh, M. Shifman and A.I. Vainshtein, 
   {\em Phys.Rev.}{\bf D49} (1994) 3356; Erratum: {\em ibid.}, {\bf D50}
   (1994) 3572.
\bibitem{mannel}T. Mannel, {\em Nucl.Phys.}{\bf B413} (1994) 396.
\bibitem{Ko} L. Koyrakh, {\em Phys.Rev.}{\bf D49} (1994) 3379.
\bibitem{balk} S. Balk, J.G. K\"orner, D. Pirjol and K. Schilcher,
   {\em Z.Phys.}{\bf C64} (1994) 37.
\bibitem{FaLi} A.F. Falk, Z. Ligeti, M. Neubert and Y. Nir,
   {\em Phys.Lett.}{\bf B326} (1994) 145.
\bibitem{CDNP} P. Colangelo, C.A. Dominguez, G. Nardulli and N. Paver,
   {\em Phys.Rev.}{\bf D54} (1996) 4622.
\bibitem{falk} A.F. Falk and M. Neubert, 
  {\em Phys.Rev.}{\bf D47} (1993) 2965.
\bibitem{koernerpirjol} J.G. K\"orner and D. Pirjol, 
   {\em Phys.Lett.}{\bf B334} (1994) 339.
\bibitem{rose} M.E. Rose, {\em Elementary theory of angular momentum},
   New York, 1957.
\bibitem{itzykson} C. Itzykson and J.B. Zuber, {\em Quantum Field Theory},
   McGraw-Hill, 1980.
\bibitem{lampe} B. Lampe, 
 {\em Nucl.Phys.}{\bf B454} (1995) 506;
 {\em ibid.} {\bf B458} (1996) 23.
\bibitem{vives} G. Barenhoim, J. Bernabeu and O. Vives, 
   {\em Phys.Rev.Lett.}{\bf 77} (1996) 3299.
\bibitem{GrLi} Y. Grossman and Z. Ligeti, {\em Phys.Lett.}{\bf B347}
   399 (1995).
\bibitem{ShVo}  M.A. Shifman and M.B. Voloshin, {\em Sov.J.Nucl.Phys.}
   {\bf 41} 120 (1985).
\end{thebibliography}
\end{document}